\documentclass[amsfonts, amssymb, amsmath, showkeys, nofootinbib, reprint,floatfix, prl, superscriptaddress]{revtex4-2}

\usepackage[colorinlistoftodos, color=green!40,prependcaption]{todonotes}
\usepackage{graphicx}
\usepackage{dcolumn}
\usepackage{bm}
\usepackage{physics}
\usepackage{float}
\usepackage{xcolor}
\usepackage{amsmath}
\usepackage{amssymb}
\usepackage{units}
\usepackage[english]{babel}
\usepackage{placeins}
\usepackage{lineno}

\usepackage{lipsum}
\usepackage{comment}

\begin{document}

\title{Remote magnon sensing detects the universal phase stiffness jump in Bi\textsubscript{2}Sr\textsubscript{2}CaCu\textsubscript{2}O\textsubscript{8+$\delta$}}
\author{Yue Sun}
\email{yuesun1@stanford.edu}
\affiliation{Department of Physics, University of Washington, Seattle, Washington 98195, USA}

\author{Mai Nguyen}
\author{Naiyuan James Zhang}
\affiliation{Department of Physics, University of Washington, Seattle, Washington 98195, USA}

\author{Dingbin Huang}
\affiliation{Materials Science Division, Lawrence Berkeley National Laboratory, Berkeley, California 94720, USA}

\author{Shuai Yuan}
\affiliation{Department of Physics, University of Washington, Seattle, Washington 98195, USA}

\author{Aljoscha S$\ddot{\textrm{o}}$ll}
\affiliation{Department of Inorganic Chemistry, University of Chemistry and Technology Prague, Technická 5, 166 28 Prague 6, Czech Republic}

\author{Arun Ramanathan}
\affiliation{Department of Chemistry, Columbia University, New York, New York 10027, USA}

\author{Takashi Taniguchi}
\affiliation{Research Center for Materials Nanoarchitectonics, National Institute for Materials Science, 1-1 Namiki, Tsukuba 305-0044, Japan}

\author{Kenji Watanabe}
\affiliation{Research Center for Electronic and Optical Materials, National Institute for Materials Science, 1-1 Namiki, Tsukuba 305-0044, Japan}

\author{Xavier Roy}
\affiliation{Department of Chemistry, Columbia University, New York, New York 10027, USA}

\author{Zden\v ek Sofer}
\affiliation{Department of Inorganic Chemistry, University of Chemistry and Technology Prague, Technická 5, 166 28 Prague 6, Czech Republic}

\author{Shigeyuki Ishida}
\author{Hiroshi Eisaki}
\affiliation{National Institute of Advanced Industrial Science and Technology (AIST), Tsukuba, Ibaraki 305-8568, Japan}

\author{Makoto Hashimoto}
\affiliation{Stanford Synchrotron Radiation Lightsource, SLAC National Accelerator Laboratory, Menlo Park, California 94025, USA}

\author{Zhi-Xun Shen}
\affiliation{Stanford Institute for Materials and Energy Sciences, SLAC National Accelerator Laboratory, Menlo Park, California 94025, USA}
\affiliation{Geballe Laboratory for Advanced Materials, Departments of Physics and Applied Physics, Stanford University, Stanford, California 94305, USA}

\author{Joseph Orenstein}
\email{jworenstein@lbl.gov}
\affiliation {Materials Science Division, Lawrence Berkeley National Laboratory, Berkeley, California 94720, USA}
\affiliation {Department of Physics, University of California, Berkeley, California 94720, USA}

\author{Xiaodong Xu}
\email{xuxd@uw.edu}
\affiliation{Department of Physics, University of Washington, Seattle, Washington 98195, USA}
\affiliation{Department of Materials Science and Engineering, University of Washington, Seattle, Washington 98195, USA}
\affiliation{Stanford Institute for Materials and Energy Sciences, SLAC National Accelerator Laboratory, Menlo Park, California 94025, USA}
\affiliation{Geballe Laboratory for Advanced Materials, Departments of Physics and Applied Physics, Stanford University, Stanford, California 94305, USA}


\begin{abstract}
Phase stiffness governs macroscopic quantum coherence in superfluids and superconductors, providing a sensitive probe of condensate depletion, fluctuations, and phase transitions. However quantitative and spatially resolved measurements remain challenging. Here we report a contactless, local, all-optical measurement of superconducting phase stiffness using propagating antiferromagnetic magnons as remote sensors. In CrSBr, strong exciton–magnon coupling enables optical excitation and detection of magnon wavepackets and precise determination of their group velocity. When CrSBr is interfaced with underdoped Bi\textsubscript{2}Sr\textsubscript{2}CaCu\textsubscript{2}O\textsubscript{8+$\delta$}, we find that the magnon group velocity is substantially enhanced below the superconducting transition and quantitatively encodes the phase stiffness through Meissner screening currents. Using this capability, we identify a parallel-field-induced Berezinskii–Kosterlitz–Thouless transition through direct observation of the universal phase-stiffness jump, revealing a field-controlled route to expose intrinsic two-dimensional phase-ordering physics in a bulk cuprate. Our work establishes propagating magnons as local probes of superconducting electrodynamics and phase-coherent phenomena in low-dimensional quantum materials.
\end{abstract}

\maketitle

The superfluid stiffness, $\rho_s$, quantifies the rigidity of a superconducting condensate against spatial variations of its phase.  The temperature dependence of $\rho_s$ reflects complementary aspects of superconductivity: its depletion by quasiparticle excitations constrains the superconducting gap structure, while its critical behavior reveals how global phase coherence is lost. In a charged superfluid, $\rho_s$ also determines the strength of Meissner screening, parameterized by the London penetration depth, $\lambda_L$. Measurements of $\lambda_L$, and hence $\rho_s(T)$, have consequently become central to investigating both microscopic pairing and collective phase phenomena in superconductors. Extending such measurements to an all-optical probe with micrometer-scale spatial resolution would open opportunities for studying local superfluid stiffness in heterostructures, buried interfaces, and 2D superconductors with strong spatial inhomogeneity.

Here, we introduce a local, contactless, and all-optical measurement of $\lambda_L$ by tracking the propagation of magnon wavepackets in the van der Waals antiferromagnet CrSBr. The strong spin-exciton coupling and long magnon coherence time in CrSBr enable localized optical excitation of spin wave packets and high-precision imaging of their propagation \cite{baeExcitoncoupledCoherentMagnons2022e,sunDipolarSpinWave2024b}. We find that placing CrSBr on an underdoped Bi\textsubscript{2}Sr\textsubscript{2}CaCu\textsubscript{2}O\textsubscript{8+$\delta$} (BSCCO) substantially enhances the magnon group velocity $v_g$, revealing that even an antiferromagnet with vanishing equilibrium net magnetization can strongly couple to Meissner screening. Our measurements show that this interaction originates from the dynamical dipolar field carried by the propagating magnon wavepacket, which induces local Meissner screening currents in the underlying superconductor and thereby renormalizes the magnon dispersion \cite{borstObservationControlHybrid2023}. Quantifying this velocity enhancement allows the extraction of London penetration depth $\lambda_L$, and thus $\rho_s$. Because excitation and detection are performed entirely through free-space optics, the method requires neither electrical contacts nor microwave structures. It remains broadly accessible across temperature and magnetic field while providing dynamical information unavailable in static standing-wave measurements.

We then use this capability to track how superconducting phase coherence evolves under a magnetic field applied parallel to the CuO$_2$ planes. Near the superconducting phase transition temperature $T_c$, we find that a smooth evolution of $\rho_s(T)$ at low fields sharpens with increasing field, reaching an abrupt jump with a relative width of $\Delta T/T_c\approx0.02$. Remarkably, the magnitude of the step agrees with the universal jump predicted for a Berezinskii–Kosterlitz–Thouless (BKT) transition to within 10\% \cite{nelsonUniversalJumpSuperfluid1977,beasleyPossibilityVortexAntivortexPair1979,bishopStudySuperfluidTransition1978}. This observation demonstrates a long-standing theoretical prediction \cite{efetovFluctuationsLayeredSuperconductors1979}: a parallel-field-induced BKT transition in bulk cuprates. 

\section{Accelerated magnon propagation by Meissner screening}
\begin{figure*}[tb]
\centering
\includegraphics[width=2.0\columnwidth]{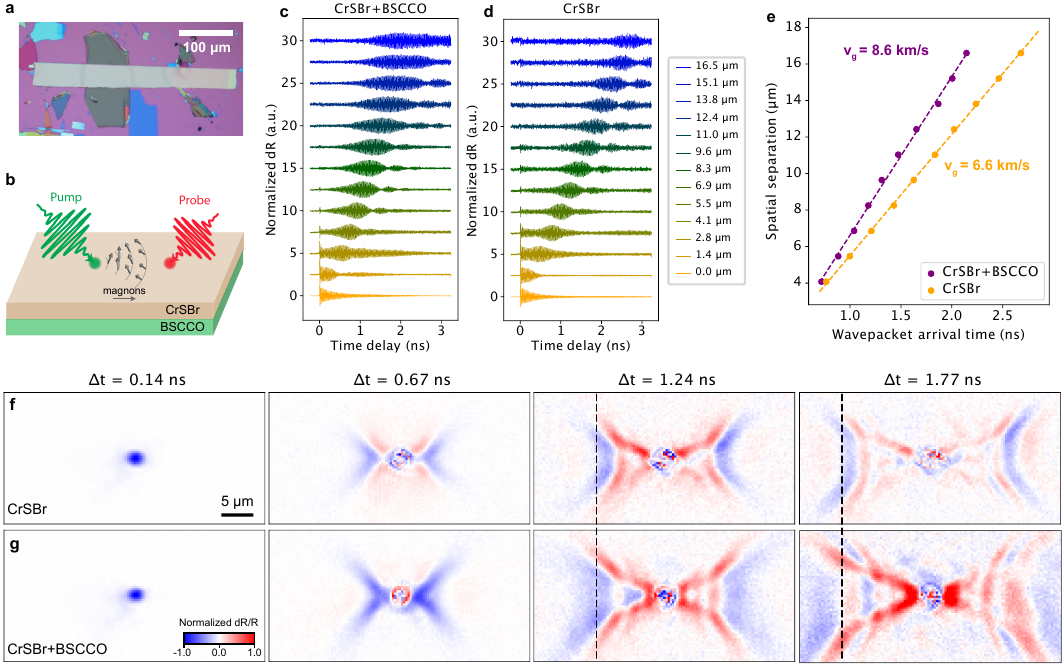}
\caption{\textbf{Accelerated magnon propagation in a CrSBr/BSCCO heterostructure}. \textbf{a}, Optical image of the CrSBr/BSCCO heterostructure. The 187 nm thick underdoped BSCCO flake has its $T_c$ at 62 K. The CrSBr flake is 385 nm thick. \textbf{b}, Schematic of the experimental setup. A pump and probe foci are spatially separated to probe the propagation of magnons on CrSBr/BSCCO heterostructure. \textbf{c},\textbf{d}, Transient reflectance measured for pump–probe separations ranging from 0.0 to 16.5 $\mu$m in (c) the heterostructure and (d) the bare CrSBr flake. \textbf{e}, Extraction of magnon group velocity $v_g$ in both CrSBr/BSCCO and bare CrSBr flake. \textbf{f},\textbf{g}, Snapshots of spatially resolved transient reflectance measured on (f) bare CrSBr flake and (g) CrSBr/BSCCO with a magnetic field of 0.3 T applied along the a-axis of CrSBr. The dashed lines mark the magnon propagation distance on bare CrSBr at the longest two time delays.}
\label{fig:Fig1_vg_demo}
\end{figure*}

The sample of interest in our study is the CrSBr/BSCCO heterostructure shown in Fig.~\ref{fig:Fig1_vg_demo}a. A 385 nm thick CrSBr crystal is placed on top of a 187 nm thick BSCCO flake that was exfoliated from a bulk underdoped sample with transition temperature $T_c=62$ K. In our initial experiments, we characterized magnon propagation with a pair of focused laser pulses (depicted in Fig.~\ref{fig:Fig1_vg_demo}b) that act as magnon source and detector. The difference in magnon propagation in CrSBr and CrSBr/BSCCO is readily measured \textit{in situ} by shifting the foci from the CrSBr regions to those with the BSCCO flake underneath. All measurements were performed with an in-plane magnetic field applied along the $a$-axis of CrSBr, the configuration in which the high-frequency magnon mode is selectively probed \cite{sunDipolarSpinWave2024b,diederichTunableInteractionExcitons2023}.

In Fig.~\ref{fig:Fig1_vg_demo}(c,d), we present time traces of transient reflectance with 13 pump-probe spatial separations up to 16.5 $\mu$m along the $a$-axis of CrSBr (horizontal direction in Fig.~\ref{fig:Fig1_vg_demo}a) for both the CrSBr/BSCCO heterostructure and the bare CrSBr flake. The measurement is performed at 3 K and with 0.3 T magnetic field. In each time trace, both the amplitude and phase of coherent magnons are captured by the oscillations of transient reflectance. For each separation, the magnon wavepacket arrives earlier in the CrSBr/BSCCO heterostructure than in the bare CrSBr flake, suggesting an enhancement in magnon group velocity, $v_g$ arising from proximity to the superconductor. To estimate the enhancement of $v_g$, we take the time at which the magnon wavepacket amplitude reaches a maximum as the arrival time. As shown in Fig.~\ref{fig:Fig1_vg_demo}e, the arrival time so defined is a linear function of the spatial separation, whose slope provides a measure of $v_g$. The slope increases from 6.6$\pm$0.1 km/s measured in CrSBr alone to 8.6$\pm$0.2 km/s in CrSBr/BSCCO corresponding to a 30\% enhancement. 

The enhanced $v_g$ can also be seen in time-resolved images of magnon propagation obtained by rastering the pump beam at fixed time delays. As shown in Fig.~\ref{fig:Fig1_vg_demo}(f,g), magnon wavepackets in both the bare CrSBr flake and the CrSBr/BSCCO heterostructure propagate primarily along the $a$-axis of CrSBr, consistent with the selective detection of the high-frequency magnon mode \cite{sunDipolarSpinWave2024b}. At the same time delay, the propagation front in CrSBr/BSCCO extends beyond the maximum distance reached in the bare CrSBr flake (black dashed lines), providing a clear visualization of the enhanced propagation velocity.

\begin{figure}[tb]
\centering
\includegraphics[width=1.0\columnwidth]{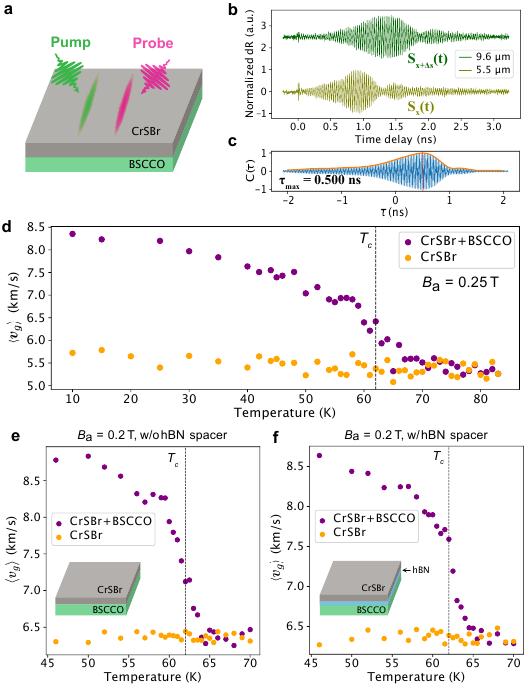}
\caption{\textbf{Correlation between $v_g$ enhancement and superconductivity}. \textbf{a}, Schematic of the experimental setup with stripe-shaped pump and probe beams. \textbf{b}, Example transient reflectance data measured for pump–probe separations at 5.50 and 9.63 $\mu$m in CrSBr/BSCCO. \textbf{c}, The correlation function $C(\tau)$ calculated from the data in (b). The envelope of $C(\tau)$ peaks at $\tau = $ 0.500 ns.  \textbf{d}, Temperature dependence of $v_g$ measured on both CrSBr/BSCCO and the CrSBr bare flake under an in-plane magnetic field of 0.25 T. \textbf{e},\textbf{f}, Comparison between the temperature dependence of $v_g$ on CrSBr/BSCCO samples (e) without and (f) with a hBN spacer. Insets show the sample geometry.}
\label{fig:Fig2_vg_temp_dep}
\end{figure}

The observations presented in Fig.~\ref{fig:Fig1_vg_demo} demonstrate that magnon propagation in CrSBr is strongly enhanced by proximity to BSCCO in its superconducting state. To elevate this demonstration to a sensitive, local probe of the absolute magnitude of $\rho_s$, we made several changes to the experimental protocol and improvements in the data analysis. First, we introduced a cylindrical lens to change the focal spots into extended line foci, reducing the dimensionality of the beam propagation from 2D to 1D and enhancing the signal reaching the detector (Fig.~\ref{fig:Fig2_vg_temp_dep}a). A second major improvement was the analysis leading to the wavepacket arrival time. Rather than tracking the maximum of individual wavepackets, we compute the correlation function of two time traces measured at a known spatial separation, $\Delta s$,
\begin{equation}
C(\tau,\Delta s) \equiv \langle S_x(t)S_{x+\Delta s}(t+\tau)\rangle_t.
\end{equation}
As presented in Fig.~\ref{fig:Fig2_vg_temp_dep}c, $C(\tau,\Delta s)$ retains the oscillatory features of the original signals, $S(t),$ but exhibits significantly improved signal-to-noise ratio, as each point in the correlation function incorporates information from the full time traces. As detailed in the Supplementary Sections 3 and 4, the group velocity averaged over the wavevector distribution of the wavepacket can be measured by the relation,
\begin{equation}
\langle v_g\rangle=\frac{\Delta s}{\tau_{max}},
\end{equation}
where $\tau_{max}$ is the arrival-time difference at which the envelope of $C(\tau,\Delta s)$ reaches its maximum. This approach is demonstrated in Fig.~\ref{fig:Fig2_vg_temp_dep}b, which shows time traces acquired at spatial separations of 5.5 $\mu$m and 9.6 $\mu$m. In this example, $C(\tau,\Delta s)$ is maximized at $\tau_{max} = 0.500$ ns, yielding 
\begin{equation*}
    \langle v_g \rangle= \frac{(9.6-5.5)\mu\text{m}}{0.500~\text{ns}}= 8.2 ~\text{km/s},
\end{equation*} 
consistent with the value obtained previously based on the arrival time of individual wavepackets. 

The improved sensitivity provided by the correlation function analysis enabled us to track the temperature dependence of $\langle v_g \rangle$ enhancement in the CrSBr/BSCCO heterostructure from low temperature through the superconducting transition.  Fig.~\ref{fig:Fig2_vg_temp_dep}d compares $\langle v_g\rangle(T)$ measured on the bare CrSBr flake (orange circles) with the CrSBr/BSCCO heterostructure (purple circles) in a 0.25 T in-plane magnetic field. The $\langle v_g\rangle(T)$ of magnon wavepackets in the bare CrSBr flake is almost constant in the temperature range of 10-83 K. In contrast, $\langle v_g\rangle(T)$ in CrSBr/BSCCO decreases with increasing $T$ and converges to the bare CrSBr velocity at approximately 5 K above $T_c$. From this we conclude that the enhancement of $\langle v_g\rangle$ in the CrSBr/BSCCO heterostructure originates from superconductivity.

The correlation between the enhancement of $\langle v_g\rangle(T)$ and the onset of superconductivity suggests two possible coupling mechanisms. One is that magnons are coupled to superconductivity via the fields arising from the Meissner current. A second possibility is that Cooper pairs in the superconductor tunnel into the antiferromagnet and alter the magnon dispersion by changing magnetic parameters. 

To distinguish these two possibilities, we prepared a sample with a 10-nm-thick hexagonal BN (hBN) spacer between the CrSBr and BSCCO flakes. The thickness of the hBN spacer is significantly larger than the coherence length of BSCCO ($\approx$ 1 nm) \cite{blatterVorticesHightemperatureSuperconductors1994}, thus preventing Cooper pairs from tunneling into the CrSBr flake. The spacer is also much thinner than CrSBr and BSCCO flakes, thus having negligible effects on dipolar fields. As shown in Figs.~\ref{fig:Fig2_vg_temp_dep}(e,f), there is no significant difference in $\langle v_g\rangle(T)$ in the two samples, excluding the possibility of coupling via Cooper pair tunneling. As we show below, a theoretical model based on magnetic dipolar coupling between magnons in CrSBr and Meissner currents in BSCCO not only explains the temperature dependence of $\langle v_g\rangle(T)$, but also enables measurement of the local phase stiffness in BSCCO.

\section{Magnon dispersion in C\MakeLowercase{r}SB\MakeLowercase{r}/BSCCO}

\begin{figure}[tb]
\centering
\includegraphics[width=1.0\columnwidth]{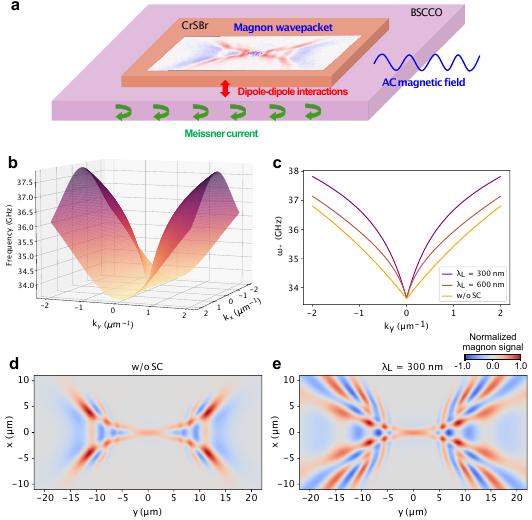}
\caption{\textbf{Magnon dispersion in antiferromagnet/superconductor heterostructures}. \textbf{a}, Cartoon illustrating the magnetic dipolar coupling between magnons in the antiferromagnet and Meissner currents in the superconductor. \textbf{b}, Calculated magnon frequency of the $\omega_+$ mode as a function of wavevector in the $k_x$-$k_y$ plane. \textbf{c}, Magnon dispersion of the $\omega_+$ mode along the $k_y$ direction for different values of $\lambda_L$. \textbf{d},\textbf{e}, Simulated spatial distribution of the $\omega_+$ magnon mode at $\Delta t=1.77$ ns (d) in a bare CrSBr flake (thickness = 385 nm) and (e) in a CrSBr/superconductor heterostructure with $\lambda_L=300$ nm.}
\label{fig:Fig3_magnon_dispersion}
\end{figure}

The strong response of antiferromagnetic magnons to superconductivity is at first glance unexpected because the equilibrium net magnetization of an antiferromagnet vanishes. Our theoretical analysis reveals, however, that the relevant coupling is governed not by the static magnetization but by the dynamical magnetic field carried by a propagating magnon wavepacket (Fig.~\ref{fig:Fig3_magnon_dispersion}a). During the magnon oscillation, the two antiferromagnetic sublattices develop a transient net magnetization that generates an AC dipolar field at the magnon frequency. This field penetrates the adjacent superconductor and induces Meissner screening currents, whose magnetic field in turn acts back on the magnons in CrSBr and renormalizes their dispersion. As shown in Supplementary Section 1, incorporating this electromagnetic coupling into the Landau-Lifshitz equation yields the dispersion relation,

 \begin{equation}
 \label{Dispersion relation}
    \omega^2_{+}(k_x,k_y) = \omega_{CSB}^2(k_x,k_y)+\frac{4\omega_0 v_0}{L}\left(\frac{k_y}{k}\right)^2\frac{\eta(k,\lambda_L,h,L)}{\lambda_L^2},
\end{equation}
where $\omega_+$ and $\omega_{CSB}$ are the magnon frequencies with, and without coupling, respectively. The second term on the right-side of Eq.~\ref{Dispersion relation} expresses the change in the magnon dispersion relation arising from Meissner screening. The parameters $\omega_0$ and $v_0$ are the frequency and group velocity of bare CrSBr evaluated at $\bm{k}=0$ and,
\begin{equation}
\begin{aligned}
   \eta(k,\lambda_L,h,L)=&\frac{\left(1-e^{-kL}\right)^2}{kL}\frac{1-e^{-2\kappa h}}{(k+\kappa)^2-(k-\kappa)^2e^{-2\kappa h}},
   \\\kappa &= \sqrt{k^2+1/\lambda_L^2}.
\end{aligned}
\end{equation}
The thickness of the BSCCO and CrSBr layers, $h$ and $L$, respectively, are determined by atomic force microscopy, while $\omega_0$ and $v_0$ are directly measured through transient reflectance from the bare CrSBr flake. The only unknown parameter is the London penetration depth, $\lambda_L$. 

The dispersion of the $\omega_{+}$ mode in the $k_x$–$k_y$ plane for $\lambda_L = 300$ nm, as shown in Fig.~\ref{fig:Fig3_magnon_dispersion}b, retains the anisotropic structure in $k$-space characteristic of the bare CrSBr flake, exhibiting strong dispersion in the $k_y$ direction and zero bandwidth in the $k_x$ direction. The dispersion for $\bm{k}=k_y\hat{y}$ is shown in Fig.~\ref{fig:Fig3_magnon_dispersion}c for several values of $\lambda_L$. Near $k = 0$, the dispersion of the $\omega_{+}$ mode becomes increasingly steep as $\lambda_L$ decreases, indicating that the group velocity enhancement observed in the CrSBr/BSCCO heterostructures reflects the emergence of superconducting order.

To test our analysis of the magnon dispersion relation in CrSBr/BSCCO heterostructures, we simulated the propagation of magnons launched by a Gaussian focal spot. We obtain the time evolution of magnon wavepackets using the relation
\begin{equation}
\bm{m}_y(x,y,t)=\bm{m}_{y}^0\int d^{2}\bm{k}\ e^{-k^{2}\sigma^{2}/2} e^{-i\omega_{+}(\bm{k})t} e^{i\bm{k}\cdot \bm{r}},
\end{equation}
where $\bm{m}_y(x,y,t)$ is the departure of the magnetization from equilibrium induced by the $\omega_{+}$ mode, $\bm{m}_{y}^{0}$ is the amplitude of the initial Gaussian excitation, $\bm{k}=(k_{x},k_{y})$, $\bm{r}=(x,y)$ and $\sigma$ is the laser spot size. Figs.~\ref{fig:Fig3_magnon_dispersion}(d,e) show the calculated spatial profiles of magnon wavepackets at $\Delta t = 1.77$ ns for a pristine CrSBr flake (Fig.~\ref{fig:Fig3_magnon_dispersion}d) and a CrSBr/BSCCO heterostructure (Fig.~\ref{fig:Fig3_magnon_dispersion}e, $\lambda_L = 300$ nm). The simulations reproduce the key features of the experimental results shown in Figs.~\ref{fig:Fig1_vg_demo}(f,g) with $\lambda_L$ as the only free parameter. 

\section{Parallel-field-induced BKT transition in BSCCO}

\begin{figure*}[tb]
\centering
\includegraphics[width=1.5\columnwidth]{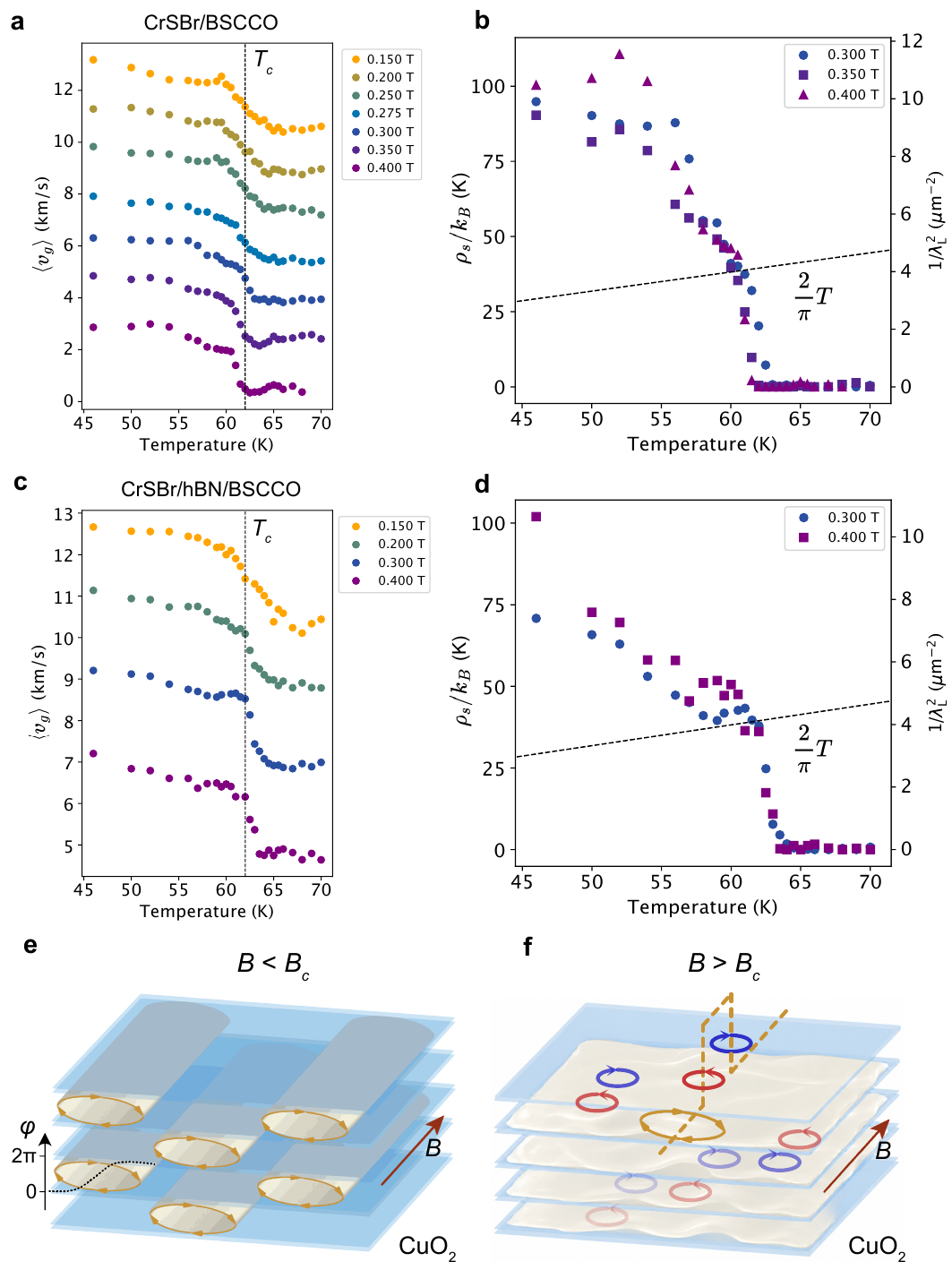}
\caption{\textbf{Observation of a parallel-field-induced BKT transition}. \textbf{a}, Temperature dependence of the averaged magnon group velocity $\langle v_g\rangle$ in CrSBr/BSCCO under in-plane magnetic fields ranging from 0.150 to 0.400 T. The magnetic field is applied along the $a$-axis of CrSBr. \textbf{b}, Extracted phase stiffness $\rho_s$ as a function of temperature. The dashed line indicates the magnitude of the universal jump predicted by the BKT theory. \textbf{c},\textbf{d}, Temperature dependence of $\langle v_g\rangle$ and extracted $\rho_s$ in the CrSBr/BSCCO device separated by an hBN spacer, demonstrating the reproducibility of the field-induced transition and the universal jump in $\rho_s$. \textbf{e}, Schematic illustration of the Josephson vortex lattice below the critical field $B_c$, where Josephson vortices remain ordered and interlayer phase coherence is preserved. This panel is adapted from Ref.~\cite{laplaceJosephsonPlasmonicsLayered2016}. \textbf{f}, Above $B_c$, overlap of the Josephson vortices give rise to strong kink--antikink fluctuations of the Josephson vortex lines, generating vortex--antivortex pairs within individual CuO$_2$ layers and an effectively two-dimensional vortex plasma.}
\label{fig:Fig4_BKT}
\end{figure*}

Having established the magnon group velocity as a quantitative probe of $\lambda_L$, a compelling application of this technique is to investigate the BKT transition in superconductors. The recognition that phase coherence in two-dimensional systems can arise from the binding of vortex–antivortex pairs brought topology to the forefront of condensed-matter physics \cite{kosterlitzOrderingMetastabilityPhase1973,berezinskiiDestructionLongrangeOrder1972,minnhagenTwodimensionalCoulombGas1987,kosterlitzNobelLectureTopological2017}. A substantial body of experimental evidence supports the emergence of BKT-like behavior in superconducting films approaching the two-dimensional limit \cite{hebardEvidenceKosterlitzThoulessTransition1980,yongEvidenceTwodimensionalQuantum2012,misraMeasurementsMagneticFieldTunedConductivity2013}. Layered superconductors such as cuprates provide a natural expression of this physics because superconductivity resides predominantly within weakly coupled CuO$_2$ planes \cite{keimerQuantumMatterHightemperature2015,basovElectrodynamicsHighSuperconductors2005,leeDopingMottInsulator2006}. However, the weak but nonzero interlayer coupling establishes three-dimensional phase coherence and preempts the transition of individual layers, blurring the emergence of BKT physics. 

A route to recovering the underlying two-dimensional physics was proposed by Efetov, who predicted that a magnetic field applied parallel to the superconducting layers could suppress interlayer phase coherence \cite{efetovFluctuationsLayeredSuperconductors1979}. Then, at sufficiently high fields, superconducting correlations would retain algebraic decay within each layer while becoming short-ranged between layers, producing an effectively decoupled two-dimensional state. A parallel magnetic field could therefore expose a BKT transition that would otherwise be preempted by three-dimensional phase coherence. However, subsequent theoretical studies have not reached a consensus on this scenario, and whether a parallel magnetic field can truly drive a BKT transition in layered superconductors, such as the cuprates, remains an open question \cite{korshunovProblemJosephsonvortexlatticeMelting1992,blatterKosterlitzThoulessTransitionSmectic1991,blatterVorticesHightemperatureSuperconductors1994}.

 Inspired by the theoretical prediction, we now investigate the effect of a magnetic field applied parallel to the CuO$_2$ layers. Fig.~\ref{fig:Fig4_BKT}a shows the temperature dependence of $\langle v_g\rangle$ under in-plane magnetic fields ranging from 0.15 T to 0.40 T. As $B_{\parallel}$ increases above $\approx 0.3$ T, the group velocity evolves from a smooth to a step-like change near $T_c$. As shown in Fig.~\ref{fig:Fig4_BKT}c, this crossover is reproduced in the heterostructure with the hBN spacer. To compare the height of the step to the BKT prediction, we first determine $\lambda_L(T)$ from $\langle v_g\rangle (T)$ using the magnon dispersion relation of the heterostructure (Eq.~\ref{Dispersion relation}). Next, the two-dimensional phase stiffness associated with each layer is given in terms of the penetation depth by the relation~\cite{beasleyPossibilityVortexAntivortexPair1979,pearlCURRENTDISTRIBUTIONSUPERCONDUCTING1964},  
\begin{equation}
\label{eq:rhos_lambdaL}
\rho_s= \frac{1}{\lambda_L^2}\frac{\hbar^2 c^2d}{16\pi e^2},
\end{equation}
where $d=$ 1.54 nm is the spacing of individual BSCCO bilayers. We emphasize that the conversion from measured $\langle v_g\rangle$ to $\lambda_L$ and finally to $\rho_s$ involves no adjustable parameters. 

By applying this conversion procedure we obtain $\rho_s(T)$ as shown in Figs.~\ref{fig:Fig4_BKT}b and \ref{fig:Fig4_BKT}d, for the CrSBr/BSCCO and CrSBr/hBN/BSCCO structures respectively. Remarkably, the jump in $\rho_s(T)$ is field-independent and quite close to the universal value $2T_{BKT}/\pi$ as indicated by the intercept of the data with the dashed lines in each of the two plots. 

Next, we consider whether the observation of the universal superfluid stiffness jump is consistent with the conjecture of field-induced layer decoupling. As shown in Fig.~\ref{fig:Fig4_BKT}e, a magnetic field applied parallel to the layers leads to the formation of  a Josephson vortex lattice (JVL) \cite{koshelevJosephsonVortexLattice2013}. The two characteristic lengths associated with the JVL are the lattice constant, $a = \Phi_0/{B_{\parallel}d}$, and the vortex core radius, $\Lambda = 2\pi\gamma d$, where $\Phi_0$ is the magnetic flux quantum and $\gamma = \lambda_c/\lambda_{ab}$ is the ratio between the London penetration depths for supercurrents flowing perpendicular to and within the CuO$_2$ planes \cite{koshelevJosephsonVortexLattice2013}. Efetov predicted that layer decoupling would onset at a field $B^\ast$ given by 
\begin{equation}
    B^\ast = \frac{\Phi_0}{2\pi\gamma d^2},
\end{equation}
which is the field at which $a = \Lambda$ and neighboring Josephson vortex cores begin to overlap  \cite{efetovFluctuationsLayeredSuperconductors1979}. With $\gamma \approx 700$ for underdoped BSCCO \cite{colsonVortexFluctuationsUnderdoped2003,blatterVorticesHightemperatureSuperconductors1994}, we obtain $B^\ast \approx 0.2$ T, in reasonable agreement with the experimentally observed crossover field. This scenario was later elaborated by Blatter, Ivlev, and Rhyner \cite{blatterKosterlitzThoulessTransitionSmectic1991}, who described the onset of decoupling as a transition from a vortex solid to a smectic vortex liquid. The low-energy excitations of these two phases are planar (or ``pancake") vortex-antivortex pairs formed when a segment of the Josephson vortex string hops to a neighboring layer (illustrated in Fig.~\ref{fig:Fig4_BKT}f). In the solid phase the string energy is a linear function of length and the pair is confined, while in the liquid phase the interaction becomes logarithmic, leading to BKT behavior. \footnote{We also notice that the parallel-field-induced layer decoupling has also been theoretically predicted to exist in double-layer quantum Hall systems through a commensurate-incommensurate transition \cite{hannaIncommensurateGroundState2001}, which may also play a role in layered superconductors.}


\section{Conclusion}

In summary, we have shown that the propagation of antiferromagnetic magnon wavepackets is a highly sensitive local probe of superconducting phase stiffness, $\rho_s$. Demonstrating this with a CrSBr/BSCCO hybrid structure, we observed a pronounced enhancement of the magnon group velocity, $v_g$, upon entering the superconducting phase. Under an in-plane magnetic field, $v_g$ exhibits a sharp step at $T_c$, reflecting a discontinuity in $\rho_s(T)$ equal to $2T_c/\pi$ that is the characteristic signature of a BKT transition.

This work establishes magnon transport as a powerful, non-invasive probe of superconducting electrodynamics and phase coherence. The ability to access phase stiffness through magnon propagation opens new opportunities to study low-dimensional superconductivity, vortex dynamics, and nonequilibrium coupling between collective excitations in hybrid quantum materials.

\section{Acknowledgements} 
X.X. thanks Eugene Demler for valuable discussions. This work was supported by the U.S. Department of Energy, Office of Basic Energy Sciences, Division of Materials Sciences and Engineering under Contract No. DE-AC02-76SF00515 and the Quantum Materials program under the Director, Office of Science, Office of Basic Energy Sciences, Materials Sciences and Engineering Division, of the U.S. Department of Energy, Contract No. DE-AC02-05CH11231. J.O received support from the Gordon and Betty Moore Foundation's EPiQS Initiative through Grant GBMF4537 to J.O. at UC Berkeley. Y.S. also acknowledges the partial support by the David J. Thouless postdoctoral fellowship at the Department of Physics, University of Washington. Z.S. was supported by the ERC-CZ programme (project no. LL2101) from the Ministry of Education Youth and Sports and by project LUAUS25268 from Ministry of Education Youth and Sports (MEYS). Synthesis work at Columbia University supported by the Air Force Office of Scientific Research (AFOSR) under MURI grant number FA9550-25-1-0288. K.W. and T.T. acknowledge support from the CREST (JPMJCR24A5), JST and World Premier International Research Center Initiative (WPI), MEXT, Japan.

\section{Author contributions}
Y.S., J.O. and X.X. conceived the project. Y.S. performed the optical measurements with assistance from D.H. M.N. fabricated all samples with conceptual and methodological input from N.J.Z. S.Y. contributed to electrical characterization of exfoliated BSCCO flakes. A.S., A.R., X.R. and Z.S. synthesized the bulk CrSBr crystals. T.T. and K.W. synthesized the hBN crystals. S.I., H.E., M.H. and Z.-X.S. provided the bulk BSCCO crystals. Theoretical analysis was performed by Y.S. and J.O. Y.S. and J.O. wrote the paper with input from X.X. All authors read the manuscript and discussed the results. 

\bibstyle{apsrev4-1}

\bibliography{Main_Text_References}

\begin{thebibliography}{26}%
\makeatletter
\providecommand \@ifxundefined [1]{%
 \@ifx{#1\undefined}
}%
\providecommand \@ifnum [1]{%
 \ifnum #1\expandafter \@firstoftwo
 \else \expandafter \@secondoftwo
 \fi
}%
\providecommand \@ifx [1]{%
 \ifx #1\expandafter \@firstoftwo
 \else \expandafter \@secondoftwo
 \fi
}%
\providecommand \natexlab [1]{#1}%
\providecommand \enquote  [1]{``#1''}%
\providecommand \bibnamefont  [1]{#1}%
\providecommand \bibfnamefont [1]{#1}%
\providecommand \citenamefont [1]{#1}%
\providecommand \href@noop [0]{\@secondoftwo}%
\providecommand \href [0]{\begingroup \@sanitize@url \@href}%
\providecommand \@href[1]{\@@startlink{#1}\@@href}%
\providecommand \@@href[1]{\endgroup#1\@@endlink}%
\providecommand \@sanitize@url [0]{\catcode `\\12\catcode `\$12\catcode `\&12\catcode `\#12\catcode `\^12\catcode `\_12\catcode `\%12\relax}%
\providecommand \@@startlink[1]{}%
\providecommand \@@endlink[0]{}%
\providecommand \url  [0]{\begingroup\@sanitize@url \@url }%
\providecommand \@url [1]{\endgroup\@href {#1}{\urlprefix }}%
\providecommand \urlprefix  [0]{URL }%
\providecommand \Eprint [0]{\href }%
\providecommand \doibase [0]{https://doi.org/}%
\providecommand \selectlanguage [0]{\@gobble}%
\providecommand \bibinfo  [0]{\@secondoftwo}%
\providecommand \bibfield  [0]{\@secondoftwo}%
\providecommand \translation [1]{[#1]}%
\providecommand \BibitemOpen [0]{}%
\providecommand \bibitemStop [0]{}%
\providecommand \bibitemNoStop [0]{.\EOS\space}%
\providecommand \EOS [0]{\spacefactor3000\relax}%
\providecommand \BibitemShut  [1]{\csname bibitem#1\endcsname}%
\let\auto@bib@innerbib\@empty
\bibitem [{\citenamefont {Bae}\ \emph {et~al.}(2022)\citenamefont {Bae}, \citenamefont {Wang}, \citenamefont {Scheie}, \citenamefont {Xu}, \citenamefont {Chica}, \citenamefont {Diederich}, \citenamefont {Cenker}, \citenamefont {Ziebel}, \citenamefont {Bai}, \citenamefont {Ren}, \citenamefont {Dean}, \citenamefont {Delor}, \citenamefont {Xu}, \citenamefont {Roy}, \citenamefont {Kent},\ and\ \citenamefont {Zhu}}]{baeExcitoncoupledCoherentMagnons2022e}%
  \BibitemOpen
  \bibfield  {author} {\bibinfo {author} {\bibfnamefont {Y.~J.}\ \bibnamefont {Bae}}, \bibinfo {author} {\bibfnamefont {J.}~\bibnamefont {Wang}}, \bibinfo {author} {\bibfnamefont {A.}~\bibnamefont {Scheie}}, \bibinfo {author} {\bibfnamefont {J.}~\bibnamefont {Xu}}, \bibinfo {author} {\bibfnamefont {D.~G.}\ \bibnamefont {Chica}}, \bibinfo {author} {\bibfnamefont {G.~M.}\ \bibnamefont {Diederich}}, \bibinfo {author} {\bibfnamefont {J.}~\bibnamefont {Cenker}}, \bibinfo {author} {\bibfnamefont {M.~E.}\ \bibnamefont {Ziebel}}, \bibinfo {author} {\bibfnamefont {Y.}~\bibnamefont {Bai}}, \bibinfo {author} {\bibfnamefont {H.}~\bibnamefont {Ren}}, \bibinfo {author} {\bibfnamefont {C.~R.}\ \bibnamefont {Dean}}, \bibinfo {author} {\bibfnamefont {M.}~\bibnamefont {Delor}}, \bibinfo {author} {\bibfnamefont {X.}~\bibnamefont {Xu}}, \bibinfo {author} {\bibfnamefont {X.}~\bibnamefont {Roy}}, \bibinfo {author} {\bibfnamefont {A.~D.}\ \bibnamefont {Kent}},\ and\ \bibinfo {author} {\bibfnamefont {X.}~\bibnamefont {Zhu}},\ }\href
  {https://doi.org/10.1038/s41586-022-05024-1} {\bibfield  {journal} {\bibinfo  {journal} {Nature}\ }\textbf {\bibinfo {volume} {609}},\ \bibinfo {pages} {282} (\bibinfo {year} {2022})}\BibitemShut {NoStop}%
\bibitem [{\citenamefont {Sun}\ \emph {et~al.}(2024)\citenamefont {Sun}, \citenamefont {Meng}, \citenamefont {Lee}, \citenamefont {Soll}, \citenamefont {Zhang}, \citenamefont {Ramesh}, \citenamefont {Yao}, \citenamefont {Sofer},\ and\ \citenamefont {Orenstein}}]{sunDipolarSpinWave2024b}%
  \BibitemOpen
  \bibfield  {author} {\bibinfo {author} {\bibfnamefont {Y.}~\bibnamefont {Sun}}, \bibinfo {author} {\bibfnamefont {F.}~\bibnamefont {Meng}}, \bibinfo {author} {\bibfnamefont {C.}~\bibnamefont {Lee}}, \bibinfo {author} {\bibfnamefont {A.}~\bibnamefont {Soll}}, \bibinfo {author} {\bibfnamefont {H.}~\bibnamefont {Zhang}}, \bibinfo {author} {\bibfnamefont {R.}~\bibnamefont {Ramesh}}, \bibinfo {author} {\bibfnamefont {J.}~\bibnamefont {Yao}}, \bibinfo {author} {\bibfnamefont {Z.}~\bibnamefont {Sofer}},\ and\ \bibinfo {author} {\bibfnamefont {J.}~\bibnamefont {Orenstein}},\ }\href {https://doi.org/10.1038/s41567-024-02387-2} {\bibfield  {journal} {\bibinfo  {journal} {Nat. Phys.}\ }\textbf {\bibinfo {volume} {20}},\ \bibinfo {pages} {794} (\bibinfo {year} {2024})}\BibitemShut {NoStop}%
\bibitem [{\citenamefont {Borst}\ \emph {et~al.}(2023)\citenamefont {Borst}, \citenamefont {Vree}, \citenamefont {Lowther}, \citenamefont {Teepe}, \citenamefont {Kurdi}, \citenamefont {Bertelli}, \citenamefont {Simon}, \citenamefont {Blanter},\ and\ \citenamefont {Van Der~Sar}}]{borstObservationControlHybrid2023}%
  \BibitemOpen
  \bibfield  {author} {\bibinfo {author} {\bibfnamefont {M.}~\bibnamefont {Borst}}, \bibinfo {author} {\bibfnamefont {P.~H.}\ \bibnamefont {Vree}}, \bibinfo {author} {\bibfnamefont {A.}~\bibnamefont {Lowther}}, \bibinfo {author} {\bibfnamefont {A.}~\bibnamefont {Teepe}}, \bibinfo {author} {\bibfnamefont {S.}~\bibnamefont {Kurdi}}, \bibinfo {author} {\bibfnamefont {I.}~\bibnamefont {Bertelli}}, \bibinfo {author} {\bibfnamefont {B.~G.}\ \bibnamefont {Simon}}, \bibinfo {author} {\bibfnamefont {Y.~M.}\ \bibnamefont {Blanter}},\ and\ \bibinfo {author} {\bibfnamefont {T.}~\bibnamefont {Van Der~Sar}},\ }\href {https://doi.org/10.1126/science.adj7576} {\bibfield  {journal} {\bibinfo  {journal} {Science}\ }\textbf {\bibinfo {volume} {382}},\ \bibinfo {pages} {430} (\bibinfo {year} {2023})}\BibitemShut {NoStop}%
\bibitem [{\citenamefont {Nelson}\ and\ \citenamefont {Kosterlitz}(1977)}]{nelsonUniversalJumpSuperfluid1977}%
  \BibitemOpen
  \bibfield  {author} {\bibinfo {author} {\bibfnamefont {D.~R.}\ \bibnamefont {Nelson}}\ and\ \bibinfo {author} {\bibfnamefont {J.~M.}\ \bibnamefont {Kosterlitz}},\ }\href {https://doi.org/10.1103/PhysRevLett.39.1201} {\bibfield  {journal} {\bibinfo  {journal} {Phys. Rev. Lett.}\ }\textbf {\bibinfo {volume} {39}},\ \bibinfo {pages} {1201} (\bibinfo {year} {1977})}\BibitemShut {NoStop}%
\bibitem [{\citenamefont {Beasley}\ \emph {et~al.}(1979)\citenamefont {Beasley}, \citenamefont {Mooij},\ and\ \citenamefont {Orlando}}]{beasleyPossibilityVortexAntivortexPair1979}%
  \BibitemOpen
  \bibfield  {author} {\bibinfo {author} {\bibfnamefont {M.~R.}\ \bibnamefont {Beasley}}, \bibinfo {author} {\bibfnamefont {J.~E.}\ \bibnamefont {Mooij}},\ and\ \bibinfo {author} {\bibfnamefont {T.~P.}\ \bibnamefont {Orlando}},\ }\href {https://doi.org/10.1103/PhysRevLett.42.1165} {\bibfield  {journal} {\bibinfo  {journal} {Phys. Rev. Lett.}\ }\textbf {\bibinfo {volume} {42}},\ \bibinfo {pages} {1165} (\bibinfo {year} {1979})}\BibitemShut {NoStop}%
\bibitem [{\citenamefont {Bishop}\ and\ \citenamefont {Reppy}(1978)}]{bishopStudySuperfluidTransition1978}%
  \BibitemOpen
  \bibfield  {author} {\bibinfo {author} {\bibfnamefont {D.~J.}\ \bibnamefont {Bishop}}\ and\ \bibinfo {author} {\bibfnamefont {J.~D.}\ \bibnamefont {Reppy}},\ }\href {https://doi.org/10.1103/PhysRevLett.40.1727} {\bibfield  {journal} {\bibinfo  {journal} {Phys. Rev. Lett.}\ }\textbf {\bibinfo {volume} {40}},\ \bibinfo {pages} {1727} (\bibinfo {year} {1978})}\BibitemShut {NoStop}%
\bibitem [{\citenamefont {Efetov}(1979)}]{efetovFluctuationsLayeredSuperconductors1979}%
  \BibitemOpen
  \bibfield  {author} {\bibinfo {author} {\bibfnamefont {K.~B.}\ \bibnamefont {Efetov}},\ }\href@noop {} {\bibfield  {journal} {\bibinfo  {journal} {Soviet Journal of Experimental and Theoretical Physics}\ }\textbf {\bibinfo {volume} {49}},\ \bibinfo {pages} {905} (\bibinfo {year} {1979})}\BibitemShut {NoStop}%
\bibitem [{\citenamefont {Diederich}\ \emph {et~al.}(2023)\citenamefont {Diederich}, \citenamefont {Cenker}, \citenamefont {Ren}, \citenamefont {Fonseca}, \citenamefont {Chica}, \citenamefont {Bae}, \citenamefont {Zhu}, \citenamefont {Roy}, \citenamefont {Cao}, \citenamefont {Xiao},\ and\ \citenamefont {Xu}}]{diederichTunableInteractionExcitons2023}%
  \BibitemOpen
  \bibfield  {author} {\bibinfo {author} {\bibfnamefont {G.~M.}\ \bibnamefont {Diederich}}, \bibinfo {author} {\bibfnamefont {J.}~\bibnamefont {Cenker}}, \bibinfo {author} {\bibfnamefont {Y.}~\bibnamefont {Ren}}, \bibinfo {author} {\bibfnamefont {J.}~\bibnamefont {Fonseca}}, \bibinfo {author} {\bibfnamefont {D.~G.}\ \bibnamefont {Chica}}, \bibinfo {author} {\bibfnamefont {Y.~J.}\ \bibnamefont {Bae}}, \bibinfo {author} {\bibfnamefont {X.}~\bibnamefont {Zhu}}, \bibinfo {author} {\bibfnamefont {X.}~\bibnamefont {Roy}}, \bibinfo {author} {\bibfnamefont {T.}~\bibnamefont {Cao}}, \bibinfo {author} {\bibfnamefont {D.}~\bibnamefont {Xiao}},\ and\ \bibinfo {author} {\bibfnamefont {X.}~\bibnamefont {Xu}},\ }\href {https://doi.org/10.1038/s41565-022-01259-1} {\bibfield  {journal} {\bibinfo  {journal} {Nat. Nanotechnol.}\ }\textbf {\bibinfo {volume} {18}},\ \bibinfo {pages} {23} (\bibinfo {year} {2023})}\BibitemShut {NoStop}%
\bibitem [{\citenamefont {Blatter}\ \emph {et~al.}(1994)\citenamefont {Blatter}, \citenamefont {Feigel'man}, \citenamefont {Geshkenbein}, \citenamefont {Larkin},\ and\ \citenamefont {Vinokur}}]{blatterVorticesHightemperatureSuperconductors1994}%
  \BibitemOpen
  \bibfield  {author} {\bibinfo {author} {\bibfnamefont {G.}~\bibnamefont {Blatter}}, \bibinfo {author} {\bibfnamefont {M.~V.}\ \bibnamefont {Feigel'man}}, \bibinfo {author} {\bibfnamefont {V.~B.}\ \bibnamefont {Geshkenbein}}, \bibinfo {author} {\bibfnamefont {A.~I.}\ \bibnamefont {Larkin}},\ and\ \bibinfo {author} {\bibfnamefont {V.~M.}\ \bibnamefont {Vinokur}},\ }\href {https://doi.org/10.1103/RevModPhys.66.1125} {\bibfield  {journal} {\bibinfo  {journal} {Rev. Mod. Phys.}\ }\textbf {\bibinfo {volume} {66}},\ \bibinfo {pages} {1125} (\bibinfo {year} {1994})}\BibitemShut {NoStop}%
\bibitem [{\citenamefont {Laplace}\ and\ \citenamefont {Cavalleri}(2016)}]{laplaceJosephsonPlasmonicsLayered2016}%
  \BibitemOpen
  \bibfield  {author} {\bibinfo {author} {\bibfnamefont {Y.}~\bibnamefont {Laplace}}\ and\ \bibinfo {author} {\bibfnamefont {A.}~\bibnamefont {Cavalleri}},\ }\href {https://doi.org/10.1080/23746149.2016.1212671} {\bibfield  {journal} {\bibinfo  {journal} {Adv. Phys. X}\ }\textbf {\bibinfo {volume} {1}},\ \bibinfo {pages} {387} (\bibinfo {year} {2016})}\BibitemShut {NoStop}%
\bibitem [{\citenamefont {Kosterlitz}\ and\ \citenamefont {Thouless}(1973)}]{kosterlitzOrderingMetastabilityPhase1973}%
  \BibitemOpen
  \bibfield  {author} {\bibinfo {author} {\bibfnamefont {J.~M.}\ \bibnamefont {Kosterlitz}}\ and\ \bibinfo {author} {\bibfnamefont {D.~J.}\ \bibnamefont {Thouless}},\ }\href {https://doi.org/10.1088/0022-3719/6/7/010} {\bibfield  {journal} {\bibinfo  {journal} {J. Phys. C: Solid State Phys.}\ }\textbf {\bibinfo {volume} {6}},\ \bibinfo {pages} {1181} (\bibinfo {year} {1973})}\BibitemShut {NoStop}%
\bibitem [{\citenamefont {Berezinski{\v i}}(1972)}]{berezinskiiDestructionLongrangeOrder1972}%
  \BibitemOpen
  \bibfield  {author} {\bibinfo {author} {\bibfnamefont {V.~L.}\ \bibnamefont {Berezinski{\v i}}},\ }\href@noop {} {\bibfield  {journal} {\bibinfo  {journal} {Soviet Journal of Experimental and Theoretical Physics}\ }\textbf {\bibinfo {volume} {34}},\ \bibinfo {pages} {610} (\bibinfo {year} {1972})}\BibitemShut {NoStop}%
\bibitem [{\citenamefont {Minnhagen}(1987)}]{minnhagenTwodimensionalCoulombGas1987}%
  \BibitemOpen
  \bibfield  {author} {\bibinfo {author} {\bibfnamefont {P.}~\bibnamefont {Minnhagen}},\ }\href {https://doi.org/10.1103/RevModPhys.59.1001} {\bibfield  {journal} {\bibinfo  {journal} {Rev. Mod. Phys.}\ }\textbf {\bibinfo {volume} {59}},\ \bibinfo {pages} {1001} (\bibinfo {year} {1987})}\BibitemShut {NoStop}%
\bibitem [{\citenamefont {Kosterlitz}(2017)}]{kosterlitzNobelLectureTopological2017}%
  \BibitemOpen
  \bibfield  {author} {\bibinfo {author} {\bibfnamefont {J.~M.}\ \bibnamefont {Kosterlitz}},\ }\href {https://doi.org/10.1103/RevModPhys.89.040501} {\bibfield  {journal} {\bibinfo  {journal} {Rev. Mod. Phys.}\ }\textbf {\bibinfo {volume} {89}},\ \bibinfo {pages} {040501} (\bibinfo {year} {2017})}\BibitemShut {NoStop}%
\bibitem [{\citenamefont {Hebard}\ and\ \citenamefont {Fiory}(1980)}]{hebardEvidenceKosterlitzThoulessTransition1980}%
  \BibitemOpen
  \bibfield  {author} {\bibinfo {author} {\bibfnamefont {A.~F.}\ \bibnamefont {Hebard}}\ and\ \bibinfo {author} {\bibfnamefont {A.~T.}\ \bibnamefont {Fiory}},\ }\href {https://doi.org/10.1103/PhysRevLett.44.291} {\bibfield  {journal} {\bibinfo  {journal} {Phys. Rev. Lett.}\ }\textbf {\bibinfo {volume} {44}},\ \bibinfo {pages} {291} (\bibinfo {year} {1980})}\BibitemShut {NoStop}%
\bibitem [{\citenamefont {Yong}\ \emph {et~al.}(2012)\citenamefont {Yong}, \citenamefont {Hinton}, \citenamefont {McCray}, \citenamefont {Randeria}, \citenamefont {Naamneh}, \citenamefont {Kanigel},\ and\ \citenamefont {Lemberger}}]{yongEvidenceTwodimensionalQuantum2012}%
  \BibitemOpen
  \bibfield  {author} {\bibinfo {author} {\bibfnamefont {J.}~\bibnamefont {Yong}}, \bibinfo {author} {\bibfnamefont {M.~J.}\ \bibnamefont {Hinton}}, \bibinfo {author} {\bibfnamefont {A.}~\bibnamefont {McCray}}, \bibinfo {author} {\bibfnamefont {M.}~\bibnamefont {Randeria}}, \bibinfo {author} {\bibfnamefont {M.}~\bibnamefont {Naamneh}}, \bibinfo {author} {\bibfnamefont {A.}~\bibnamefont {Kanigel}},\ and\ \bibinfo {author} {\bibfnamefont {T.~R.}\ \bibnamefont {Lemberger}},\ }\href {https://doi.org/10.1103/PhysRevB.85.180507} {\bibfield  {journal} {\bibinfo  {journal} {Phys. Rev. B}\ }\textbf {\bibinfo {volume} {85}},\ \bibinfo {pages} {180507} (\bibinfo {year} {2012})}\BibitemShut {NoStop}%
\bibitem [{\citenamefont {Misra}\ \emph {et~al.}(2013)\citenamefont {Misra}, \citenamefont {Urban}, \citenamefont {Kim}, \citenamefont {Sambandamurthy},\ and\ \citenamefont {Yazdani}}]{misraMeasurementsMagneticFieldTunedConductivity2013}%
  \BibitemOpen
  \bibfield  {author} {\bibinfo {author} {\bibfnamefont {S.}~\bibnamefont {Misra}}, \bibinfo {author} {\bibfnamefont {L.}~\bibnamefont {Urban}}, \bibinfo {author} {\bibfnamefont {M.}~\bibnamefont {Kim}}, \bibinfo {author} {\bibfnamefont {G.}~\bibnamefont {Sambandamurthy}},\ and\ \bibinfo {author} {\bibfnamefont {A.}~\bibnamefont {Yazdani}},\ }\href {https://doi.org/10.1103/PhysRevLett.110.037002} {\bibfield  {journal} {\bibinfo  {journal} {Phys. Rev. Lett.}\ }\textbf {\bibinfo {volume} {110}},\ \bibinfo {pages} {037002} (\bibinfo {year} {2013})}\BibitemShut {NoStop}%
\bibitem [{\citenamefont {Keimer}\ \emph {et~al.}(2015)\citenamefont {Keimer}, \citenamefont {Kivelson}, \citenamefont {Norman}, \citenamefont {Uchida},\ and\ \citenamefont {Zaanen}}]{keimerQuantumMatterHightemperature2015}%
  \BibitemOpen
  \bibfield  {author} {\bibinfo {author} {\bibfnamefont {B.}~\bibnamefont {Keimer}}, \bibinfo {author} {\bibfnamefont {S.~A.}\ \bibnamefont {Kivelson}}, \bibinfo {author} {\bibfnamefont {M.~R.}\ \bibnamefont {Norman}}, \bibinfo {author} {\bibfnamefont {S.}~\bibnamefont {Uchida}},\ and\ \bibinfo {author} {\bibfnamefont {J.}~\bibnamefont {Zaanen}},\ }\href {https://doi.org/10.1038/nature14165} {\bibfield  {journal} {\bibinfo  {journal} {Nature}\ }\textbf {\bibinfo {volume} {518}},\ \bibinfo {pages} {179} (\bibinfo {year} {2015})}\BibitemShut {NoStop}%
\bibitem [{\citenamefont {Basov}\ and\ \citenamefont {Timusk}(2005)}]{basovElectrodynamicsHighSuperconductors2005}%
  \BibitemOpen
  \bibfield  {author} {\bibinfo {author} {\bibfnamefont {D.~N.}\ \bibnamefont {Basov}}\ and\ \bibinfo {author} {\bibfnamefont {T.}~\bibnamefont {Timusk}},\ }\href {https://doi.org/10.1103/RevModPhys.77.721} {\bibfield  {journal} {\bibinfo  {journal} {Rev. Mod. Phys.}\ }\textbf {\bibinfo {volume} {77}},\ \bibinfo {pages} {721} (\bibinfo {year} {2005})}\BibitemShut {NoStop}%
\bibitem [{\citenamefont {Lee}\ \emph {et~al.}(2006)\citenamefont {Lee}, \citenamefont {Nagaosa},\ and\ \citenamefont {Wen}}]{leeDopingMottInsulator2006}%
  \BibitemOpen
  \bibfield  {author} {\bibinfo {author} {\bibfnamefont {P.~A.}\ \bibnamefont {Lee}}, \bibinfo {author} {\bibfnamefont {N.}~\bibnamefont {Nagaosa}},\ and\ \bibinfo {author} {\bibfnamefont {X.-G.}\ \bibnamefont {Wen}},\ }\href {https://doi.org/10.1103/RevModPhys.78.17} {\bibfield  {journal} {\bibinfo  {journal} {Rev. Mod. Phys.}\ }\textbf {\bibinfo {volume} {78}},\ \bibinfo {pages} {17} (\bibinfo {year} {2006})}\BibitemShut {NoStop}%
\bibitem [{\citenamefont {Korshunov}\ and\ \citenamefont {Larkin}(1992)}]{korshunovProblemJosephsonvortexlatticeMelting1992}%
  \BibitemOpen
  \bibfield  {author} {\bibinfo {author} {\bibfnamefont {S.~E.}\ \bibnamefont {Korshunov}}\ and\ \bibinfo {author} {\bibfnamefont {A.~I.}\ \bibnamefont {Larkin}},\ }\href {https://doi.org/10.1103/PhysRevB.46.6395} {\bibfield  {journal} {\bibinfo  {journal} {Phys. Rev. B}\ }\textbf {\bibinfo {volume} {46}},\ \bibinfo {pages} {6395} (\bibinfo {year} {1992})}\BibitemShut {NoStop}%
\bibitem [{\citenamefont {Blatter}\ \emph {et~al.}(1991)\citenamefont {Blatter}, \citenamefont {Ivlev},\ and\ \citenamefont {Rhyner}}]{blatterKosterlitzThoulessTransitionSmectic1991}%
  \BibitemOpen
  \bibfield  {author} {\bibinfo {author} {\bibfnamefont {G.}~\bibnamefont {Blatter}}, \bibinfo {author} {\bibfnamefont {B.~I.}\ \bibnamefont {Ivlev}},\ and\ \bibinfo {author} {\bibfnamefont {J.}~\bibnamefont {Rhyner}},\ }\href {https://doi.org/10.1103/PhysRevLett.66.2392} {\bibfield  {journal} {\bibinfo  {journal} {Phys. Rev. Lett.}\ }\textbf {\bibinfo {volume} {66}},\ \bibinfo {pages} {2392} (\bibinfo {year} {1991})}\BibitemShut {NoStop}%
\bibitem [{\citenamefont {Pearl}(1964)}]{pearlCURRENTDISTRIBUTIONSUPERCONDUCTING1964}%
  \BibitemOpen
  \bibfield  {author} {\bibinfo {author} {\bibfnamefont {J.}~\bibnamefont {Pearl}},\ }\href {https://doi.org/10.1063/1.1754056} {\bibfield  {journal} {\bibinfo  {journal} {Appl. Phys. Lett.}\ }\textbf {\bibinfo {volume} {5}},\ \bibinfo {pages} {65} (\bibinfo {year} {1964})}\BibitemShut {NoStop}%
\bibitem [{\citenamefont {Koshelev}\ and\ \citenamefont {Dodgson}(2013)}]{koshelevJosephsonVortexLattice2013}%
  \BibitemOpen
  \bibfield  {author} {\bibinfo {author} {\bibfnamefont {A.~E.}\ \bibnamefont {Koshelev}}\ and\ \bibinfo {author} {\bibfnamefont {M.~J.~W.}\ \bibnamefont {Dodgson}},\ }\href {https://doi.org/10.1134/S1063776113110125} {\bibfield  {journal} {\bibinfo  {journal} {Soviet Journal of Experimental and Theoretical Physics}\ }\textbf {\bibinfo {volume} {117}},\ \bibinfo {pages} {449} (\bibinfo {year} {2013})}\BibitemShut {NoStop}%
\bibitem [{\citenamefont {Colson}\ \emph {et~al.}(2003)\citenamefont {Colson}, \citenamefont {Konczykowski}, \citenamefont {Gaifullin}, \citenamefont {Matsuda}, \citenamefont {Gier{\l}owski}, \citenamefont {Li}, \citenamefont {Kes},\ and\ \citenamefont {Van Der~Beek}}]{colsonVortexFluctuationsUnderdoped2003}%
  \BibitemOpen
  \bibfield  {author} {\bibinfo {author} {\bibfnamefont {S.}~\bibnamefont {Colson}}, \bibinfo {author} {\bibfnamefont {M.}~\bibnamefont {Konczykowski}}, \bibinfo {author} {\bibfnamefont {M.~B.}\ \bibnamefont {Gaifullin}}, \bibinfo {author} {\bibfnamefont {Y.}~\bibnamefont {Matsuda}}, \bibinfo {author} {\bibfnamefont {P.}~\bibnamefont {Gier{\l}owski}}, \bibinfo {author} {\bibfnamefont {M.}~\bibnamefont {Li}}, \bibinfo {author} {\bibfnamefont {P.~H.}\ \bibnamefont {Kes}},\ and\ \bibinfo {author} {\bibfnamefont {C.~J.}\ \bibnamefont {Van Der~Beek}},\ }\href {https://doi.org/10.1103/PhysRevLett.90.137002} {\bibfield  {journal} {\bibinfo  {journal} {Phys. Rev. Lett.}\ }\textbf {\bibinfo {volume} {90}},\ \bibinfo {pages} {137002} (\bibinfo {year} {2003})}\BibitemShut {NoStop}%
\bibitem [{\citenamefont {Hanna}\ \emph {et~al.}(2001)\citenamefont {Hanna}, \citenamefont {MacDonald},\ and\ \citenamefont {Girvin}}]{hannaIncommensurateGroundState2001}%
  \BibitemOpen
  \bibfield  {author} {\bibinfo {author} {\bibfnamefont {C.~B.}\ \bibnamefont {Hanna}}, \bibinfo {author} {\bibfnamefont {A.~H.}\ \bibnamefont {MacDonald}},\ and\ \bibinfo {author} {\bibfnamefont {S.~M.}\ \bibnamefont {Girvin}},\ }\href {https://doi.org/10.1103/PhysRevB.63.125305} {\bibfield  {journal} {\bibinfo  {journal} {Phys. Rev. B}\ }\textbf {\bibinfo {volume} {63}},\ \bibinfo {pages} {125305} (\bibinfo {year} {2001})}\BibitemShut {NoStop}%
\end{thebibliography}%


\begin{thebibliography}{4}%
\makeatletter
\providecommand \@ifxundefined [1]{%
 \@ifx{#1\undefined}
}%
\providecommand \@ifnum [1]{%
 \ifnum #1\expandafter \@firstoftwo
 \else \expandafter \@secondoftwo
 \fi
}%
\providecommand \@ifx [1]{%
 \ifx #1\expandafter \@firstoftwo
 \else \expandafter \@secondoftwo
 \fi
}%
\providecommand \natexlab [1]{#1}%
\providecommand \enquote  [1]{``#1''}%
\providecommand \bibnamefont  [1]{#1}%
\providecommand \bibfnamefont [1]{#1}%
\providecommand \citenamefont [1]{#1}%
\providecommand \href@noop [0]{\@secondoftwo}%
\providecommand \href [0]{\begingroup \@sanitize@url \@href}%
\providecommand \@href[1]{\@@startlink{#1}\@@href}%
\providecommand \@@href[1]{\endgroup#1\@@endlink}%
\providecommand \@sanitize@url [0]{\catcode `\\12\catcode `\$12\catcode `\&12\catcode `\#12\catcode `\^12\catcode `\_12\catcode `\%12\relax}%
\providecommand \@@startlink[1]{}%
\providecommand \@@endlink[0]{}%
\providecommand \url  [0]{\begingroup\@sanitize@url \@url }%
\providecommand \@url [1]{\endgroup\@href {#1}{\urlprefix }}%
\providecommand \urlprefix  [0]{URL }%
\providecommand \Eprint [0]{\href }%
\providecommand \doibase [0]{https://doi.org/}%
\providecommand \selectlanguage [0]{\@gobble}%
\providecommand \bibinfo  [0]{\@secondoftwo}%
\providecommand \bibfield  [0]{\@secondoftwo}%
\providecommand \translation [1]{[#1]}%
\providecommand \BibitemOpen [0]{}%
\providecommand \bibitemStop [0]{}%
\providecommand \bibitemNoStop [0]{.\EOS\space}%
\providecommand \EOS [0]{\spacefactor3000\relax}%
\providecommand \BibitemShut  [1]{\csname bibitem#1\endcsname}%
\let\auto@bib@innerbib\@empty
\bibitem [{\citenamefont {Kalinikos}\ and\ \citenamefont {Slavin}(1986)}]{kalinikosTheoryDipoleexchangeSpin1986a}%
  \BibitemOpen
  \bibfield  {author} {\bibinfo {author} {\bibfnamefont {B.~A.}\ \bibnamefont {Kalinikos}}\ and\ \bibinfo {author} {\bibfnamefont {A.~N.}\ \bibnamefont {Slavin}},\ }\href {https://doi.org/10.1088/0022-3719/19/35/014} {\bibfield  {journal} {\bibinfo  {journal} {Journal of Physics C Solid State Physics}\ }\textbf {\bibinfo {volume} {19}},\ \bibinfo {pages} {7013} (\bibinfo {year} {1986})}\BibitemShut {NoStop}%
\bibitem [{\citenamefont {Matsumoto}\ and\ \citenamefont {Murakami}(2011{\natexlab{a}})}]{matsumotoRotationalMotionMagnons2011}%
  \BibitemOpen
  \bibfield  {author} {\bibinfo {author} {\bibfnamefont {R.}~\bibnamefont {Matsumoto}}\ and\ \bibinfo {author} {\bibfnamefont {S.}~\bibnamefont {Murakami}},\ }\href {https://doi.org/10.1103/PhysRevB.84.184406} {\bibfield  {journal} {\bibinfo  {journal} {Phys. Rev. B}\ }\textbf {\bibinfo {volume} {84}},\ \bibinfo {pages} {184406} (\bibinfo {year} {2011}{\natexlab{a}})}\BibitemShut {NoStop}%
\bibitem [{\citenamefont {Matsumoto}\ and\ \citenamefont {Murakami}(2011{\natexlab{b}})}]{matsumotoTheoreticalPredictionRotating2011}%
  \BibitemOpen
  \bibfield  {author} {\bibinfo {author} {\bibfnamefont {R.}~\bibnamefont {Matsumoto}}\ and\ \bibinfo {author} {\bibfnamefont {S.}~\bibnamefont {Murakami}},\ }\href {https://doi.org/10.1103/PhysRevLett.106.197202} {\bibfield  {journal} {\bibinfo  {journal} {Phys. Rev. Lett.}\ }\textbf {\bibinfo {volume} {106}},\ \bibinfo {pages} {197202} (\bibinfo {year} {2011}{\natexlab{b}})}\BibitemShut {NoStop}%
\bibitem [{\citenamefont {Borst}\ \emph {et~al.}(2023)\citenamefont {Borst}, \citenamefont {Vree}, \citenamefont {Lowther}, \citenamefont {Teepe}, \citenamefont {Kurdi}, \citenamefont {Bertelli}, \citenamefont {Simon}, \citenamefont {Blanter},\ and\ \citenamefont {Van Der~Sar}}]{borstObservationControlHybrid2023}%
  \BibitemOpen
  \bibfield  {author} {\bibinfo {author} {\bibfnamefont {M.}~\bibnamefont {Borst}}, \bibinfo {author} {\bibfnamefont {P.~H.}\ \bibnamefont {Vree}}, \bibinfo {author} {\bibfnamefont {A.}~\bibnamefont {Lowther}}, \bibinfo {author} {\bibfnamefont {A.}~\bibnamefont {Teepe}}, \bibinfo {author} {\bibfnamefont {S.}~\bibnamefont {Kurdi}}, \bibinfo {author} {\bibfnamefont {I.}~\bibnamefont {Bertelli}}, \bibinfo {author} {\bibfnamefont {B.~G.}\ \bibnamefont {Simon}}, \bibinfo {author} {\bibfnamefont {Y.~M.}\ \bibnamefont {Blanter}},\ and\ \bibinfo {author} {\bibfnamefont {T.}~\bibnamefont {Van Der~Sar}},\ }\href {https://doi.org/10.1126/science.adj7576} {\bibfield  {journal} {\bibinfo  {journal} {Science}\ }\textbf {\bibinfo {volume} {382}},\ \bibinfo {pages} {430} (\bibinfo {year} {2023})}\BibitemShut {NoStop}%
\end{thebibliography}%

\end{document}


\title{Remote magnon sensing detects the universal phase stiffness jump in Bi\textsubscript{2}Sr\textsubscript{2}CaCu\textsubscript{2}O\textsubscript{8+$\delta$}: supplementary information}
\maketitle
\tableofcontents
\newpage
\section{Dipolar magnon dispersion in an antiferromagnet/superconductor heterostructure}
The section calculates how superconductors change magnon dispersion of an antiferromagnet stacked above it. In this section, we will show that the imaginary part of optical conductivity for superconductors changes the magnon dispersion and the real part contributes an additional damping to magnons. Gaussian units are used throughout this section.

\subsection{Background: Dipolar tensor}
We first introduce the dipolar tensor which directly connects magnetization and its stray (dipolar) magnetic field. This tensor has been widely used in many theory references \cite{kalinikosTheoryDipoleexchangeSpin1986a,matsumotoRotationalMotionMagnons2011,matsumotoTheoreticalPredictionRotating2011}. This section provides the derivation of the dipolar tensor. If readers are familiar with the tensor, they can directly go to Eq.~\ref{eq:dipolar_tensor}.

In the magnetostatics limit, we have
\begin{equation}
\label{eq:div_B}
    \nabla\cdot \bm{B}=0,
\end{equation}
\begin{equation}
\label{eq:curl_H}
    \nabla\times \bm{H}=0.
\end{equation}
Eq.~\ref{eq:curl_H} allows us to define a scalar potential $\Phi$ such that
\begin{equation}
\label{eq:def_pot}
    \bm{H} = -\nabla\Phi.
\end{equation}
Considering $\bm{B} = \bm{H} + 4\pi\bm{M}$, Eq.~\ref{eq:div_B} gives
\begin{equation}
\label{eq:phi}
    \begin{aligned}
    &\nabla\cdot\bm{H}=-4\pi\nabla\cdot\bm{M},
    \\&\nabla^2\Phi=4\pi\nabla\cdot\bm{M}.
    \end{aligned}
\end{equation}
Eq.~\ref{eq:phi} can be solved via the Green's function $G(\bm{r},\bm{r}')$.
\begin{equation}
\label{eq:poisson}
    \nabla^2G(\bm{r},\bm{r}')=\delta(\bm{r}-\bm{r}')\ \to\ G(\bm{r},\bm{r}')=-\frac{1}{4\pi|\bm{r}-\bm{r}'|}.
\end{equation}
\begin{equation}
\Phi(\bm{r})=-\int d\bm{r}'\frac{\nabla_{\bm{r}'}\cdot\bm{M}(\bm{r}')}{|\bm{r}-\bm{r}'|}.
\end{equation}
Using the relation
\begin{equation}
    \nabla_{\bm{r}'}\cdot\left(\frac{\bm{M}(\bm{r}')}{|\bm{r}-\bm{r}'|}\right) = \frac{\nabla_{\bm{r}'}\cdot\bm{M}(\bm{r}')}{|\bm{r}-\bm{r}'|}+\bm{M}(\bm{r}')\cdot\nabla_{\bm{r}'}\frac{1}{|\bm{r}-\bm{r}'|},
\end{equation}
We have
\begin{equation}
    \Phi(\bm{r})=-\int d\bm{r}'\nabla_{\bm{r}'}\cdot\left(\frac{\bm{M}(\bm{r}')}{|\bm{r}-\bm{r}'|}\right)+\int d\bm{r}' \bm{M}(\bm{r}')\cdot\nabla_{\bm{r}'}\frac{1}{|\bm{r}-\bm{r}'|}
\end{equation}
Using the divergence theorem, the first term is
\begin{equation}
\label{eq:surface_integral}
    \int d\bm{r}'\nabla_{\bm{r}'}\cdot\left(\frac{\bm{M}(\bm{r}')}{|\bm{r}-\bm{r}'|}\right)=\oiint\limits_{S} \frac{\hat{\bm{n}}\cdot\bm{M}(\bm{r}')}{|\bm{r}-\bm{r}'|} dS.
\end{equation}
By choosing a surface extending to infinity, we can show that Eq.~\ref{eq:surface_integral} is equal to zero. Now we have
\begin{equation}
\label{eq:scalar_pot}
    \Phi(\bm{r}) = \int d\bm{r}' \bm{M}(\bm{r}')\cdot\nabla_{\bm{r}'}\frac{1}{|\bm{r}-\bm{r}'|}.
\end{equation}
Substituting Eq.~\ref{eq:scalar_pot} into Eq.~\ref{eq:def_pot}, we got the relation between $\bm{M}$ and $\bm{H}$.
\begin{equation}
\begin{aligned}
    \bm{H}(\bm{r}) &= -\int d\bm{r}'\bm{M}(\bm{r}')\cdot\left(\nabla_{\bm{r}}\nabla_{\bm{r}}'\frac{1}{|\bm{r}-\bm{r}'|}\right)
    \\&\equiv -\int d\bm{r}'\Gamma(\bm{r},\bm{r}')\bm{M}(\bm{r}'),
\end{aligned}
\end{equation}
where
\begin{equation}
    \Gamma(\bm{r},\bm{r}') = \nabla_{\bm{r}}\nabla_{\bm{r}'}\frac{1}{|\bm{r}-\bm{r}'|}
\end{equation}
is the dipolar tensor. In our measurements, we study thin flakes with large lateral dimensions but small thickness, so we would like to obtain the dipolar tensor in the limit of infinite lateral dimensions and finite thickness. To achieve this goal, we perform Fourier transform to Eq.~\ref{eq:poisson} with respect to the two lateral dimensions $x,y$ and define an in-plane wavevector $\bm{k}=(k_x,k_y)$.
\begin{equation}
    \left(\frac{\partial^2}{\partial z^2}-k^2\right)G(\bm{k},z-z')=\delta(z-z') \to G(\bm{k},z-z') = -\frac{1}{2k}e^{-k|z-z'|}.
\end{equation}
Considering $G(\bm{r},\bm{r}') = \int \frac{d\bm{k}}{(2\pi)^2}G(\bm{k},z-z')e^{i\bm{k}\cdot(\bm{\rho}-\bm{\rho}')}$, we have
\begin{equation}
\label{eq:coulomb_kernel}
    \frac{1}{|\bm{r}-\bm{r}'|}=\int \frac{d\bm{k}}{2\pi}\frac{1}{k}e^{-k|z-z'|}e^{i\bm{k}\cdot(\bm{\rho}-\bm{\rho}')}.
\end{equation}
Using the identity in Eq.~\ref{eq:coulomb_kernel}, we can calculate the tensor $\Gamma(\bm{k},z-z')$. Defining $\alpha,\beta=x$ or $y$,
\begin{equation}
\begin{aligned}
    \Gamma_{\alpha\beta}(\bm{r}-\bm{r}')&=\frac{\partial}{\partial r_{\alpha}}\frac{\partial}{\partial r_{\beta}'}\frac{1}{|\bm{r}-\bm{r}'|}
    \\&=\int\frac{d\bm{k}}{2\pi}\frac{k_{\alpha}k_{\beta}}{k}e^{-k|z-z'|}e^{i\bm{k}\cdot(\bm{\rho}-\bm{\rho}')}
    \\&=\int\frac{d\bm{k}}{(2\pi)^2}\Gamma_{\alpha\beta}(\bm{k},z-z')e^{i\bm{k}\cdot(\bm{\rho}-\bm{\rho}')}.
\end{aligned}
\end{equation}
Therefore,
\begin{equation}
    \Gamma_{\alpha\beta}(\bm{k},z-z')=2\pi\frac{k_\alpha k_\beta}{k}e^{-k|z-z'|}.
\end{equation}
Following similar procedures, we have
\begin{equation}
    \Gamma_{z\alpha}(\bm{k},z-z')=\Gamma_{\alpha z}(\bm{k},z-z')=2\pi i\sign(z-z')k_\alpha e^{-k|z-z'|}.
\end{equation}
The only term left now is $\Gamma_{zz}(\bm{k},z-z')$. Note 
\begin{equation}
\Tr[\Gamma(\bm{r},\bm{r}')] = \Tr\left[\nabla_{\bm{r}}\nabla_{\bm{r}'}\frac{1}{|\bm{r}-\bm{r}'|}\right]=\nabla^2\frac{1}{|\bm{r}-\bm{r}'|}=4\pi\nabla^2 G(\bm{r},\bm{r}')=4\pi\delta(\bm{r}-\bm{r}'),
\end{equation} 
\begin{equation}
\Tr[\Gamma(\bm{k},z-z')] = 4\pi\delta(z-z').
\end{equation}
We have
\begin{equation}
\begin{aligned}
    \Gamma_{zz}(\bm{k},z-z') &= \Tr[\Gamma(\bm{k},z-z')]-\Gamma_{xx}(\bm{k},z-z')-\Gamma_{yy}(\bm{k},z-z')
    \\&=4\pi\delta(z-z')-2\pi ke^{-k|z-z'|}.
\end{aligned}
\end{equation}
In summary, the dipolar tensor is
\begin{equation}
\label{eq:dipolar_tensor}
    \Gamma(\bm{k},z-z')=2\pi k\begin{bmatrix}
\left(\frac{k_x}{k}\right)^2e^{-k|z-z'|} & \frac{k_xk_y}{k^2}e^{-k|z-z'|} & i\sign(z-z')\frac{k_x}{k}e^{-k|z-z'|} \\
\frac{k_xk_y}{k^2}e^{-k|z-z'|} & \left(\frac{k_y}{k}\right)^2e^{-k|z-z'|} & i\sign(z-z')\frac{k_y}{k}e^{-k|z-z'|} \\
i\sign(z-z')\frac{k_x}{k}e^{-k|z-z'|} & i\sign(z-z')\frac{k_y}{k}e^{-k|z-z'|} & \frac{2\delta(z-z')}{k}-e^{-k|z-z'|} \\
    \end{bmatrix},
\end{equation}
where $\bm{k} = (k_x,k_y)$ and $k=|\bm{k}|=\sqrt{k_x^2+k_y^2}$.

\subsection{Dipolar magnon dispersion of antiferromagnets without superconductors}
In this section, we will set a framework for calculating the dipolar magnon dispersion of antiferromagnets with respect to the net transient magnetization $\bm{m}$. This framework is convenient for further calculations because dynamical dipolar fields are only dependent on $\bm{m}$.

\subsubsection{Total Energy}
We start by evaluating the total energy of the antiferromagnetic system with magnons. There are three components in the total energy,
\begin{equation}
E = E_{ani} + E_{ex} + E_{stat},
\end{equation}
where $E_{ani}$ is the anisotropy energy, $E_{ex}$ is the exchange energy and $E_{stat}$ is the magnetostatic energy. Assume the easy axis is along the x axis, the magnetizations of two sublattices are
\begin{equation}
\begin{aligned}
&\bm{M}_{1} = (M_{1x},\ m_{1y},\ m_{1z}),
\\&\bm{M}_{2} = (M_{2x},\ m_{2y},\ m_{2z}),
\\& M_{1x} = \sqrt{M_{s}^{2} - m_{1y}^{2} - m_{1z}^{2}},
\\& M_{2x} = -\sqrt{M_{s}^{2} - m_{2y}^{2} - m_{2z}^{2}},
\\& m_{1y},\ m_{1z},\ m_{2y},\ m_{2z} \ll M_{s}.
\end{aligned}
\end{equation}
This definition constrains the saturation magnetization of each sublattice to be $M_{s}$.

\subsubsection{Anisotropy energy}
To model CrSBr, we have a hard-axis anisotropy along $z$-axis and an easy-axis anisotropy along $x$-axis.
\begin{equation}
\begin{aligned}
E_{ani} &= -\frac{K_{x}}{M_{s}^{2}}\left(M_{1x}^{2}+M_{2x}^{2}\right) + \frac{K_{z}}{M_{s}^{2}}\left(m_{1z}^{2}+m_{2z}^{2}\right)
\\&
= -\frac{K_{x}}{M_{s}^{2}}\left(2M_{s}^{2}-m_{1y}^{2}-m_{1z}^{2}-m_{2y}^{2}-m_{2z}^{2}\right) + \frac{K_{z}}{M_{s}^{2}}\left(m_{1z}^{2}+m_{2z}^{2}\right)
\\&=-2K_{x} + \frac{K_{x}}{M_{s}^{2}}\left(m_{1y}^{2}+m_{2y}^{2}\right) + \frac{K_{x}+K_{z}}{M_{s}^{2}}\left(m_{1z}^{2} + m_{2z}^{2}\right).
\end{aligned}
\end{equation}
The anisotropy energy can be written with respect to $m_{y} = m_{1y} + m_{2y}$, $m_{z} = m_{1z} + m_{2z}$, $l_{y} = m_{1y} - m_{2y}$ and $l_{z} = m_{1z} - m_{2z}$.
\begin{equation}
E_{ani}=-2K_{x} + \frac{K_{x}}{2M_{s}^{2}}\left(m_{y}^{2}+l_{y}^{2}\right) + \frac{K_{T}}{2M_{s}^{2}}\left(m_{z}^{2} + l_{z}^{2}\right),
\end{equation}
where $K_{T} = K_{x} + K_{z}$.

\subsubsection{Exchange energy}
\begin{equation}
\begin{aligned}
E_{ex} &= \frac{J}{M_{s}^{2}}\bm{M}_{1}\cdot\bm{M}_{2}
\\&= \frac{J}{M_{s}^{2}}\left(-\sqrt{\left(M_{s}^{2} - m_{1y}^{2} - m_{1z}^{2}\right)\left(M_{s}^{2} - m_{2y}^{2} - m_{2z}^{2}\right)}+m_{1y}m_{2y}+m_{1z}m_{2z}\right).
\end{aligned}
\end{equation}
Expand to the second order of $m_{1y}$, $m_{1z}$, $m_{2y}$ and $m_{2z}$,
\begin{equation}
\begin{aligned}
E_{ex} &= -J + \frac{J}{2M_{s}^{2}}\left(m_{1y}^{2}+m_{2y}^{2}+m_{1z}^{2}+m_{2z}^{2}+2m_{1y}m_{2y}+2m_{1z}m_{2z}\right)
\\&= -J + \frac{J}{2M_{s}^{2}}(m_{y}^{2}+m_{z}^{2}).
\end{aligned}
\end{equation}

\subsubsection{Magnetostatic energy}
Since antiferromagnets have zero net magnetization in zero field, the magnetostatic energy is purely contributed by the transient magnetization $m_y$ and $m_z$. In this note, we are only interested in the dipolar magnon mode with the lowest order $z$-dependence, so we assume $m_y$ and $m_z$ are uniform along the $z$-direction. In an antiferromagnet with a thickness of $L$, the dipolar field $\bm{H}_D(\bm{k})$ is
\begin{equation}
    \bm{H}_D(\bm{k})=\frac{1}{L}\int_0^Ldz \bm{H}_D(\bm{k},z)=-\frac{1}{L}\int_0^Ldz\int_0^Ldz'\Gamma(\bm{k},z-z')\bm{m},
\end{equation}
where $\bm{m}=(0,m_y,m_z)$. Using the integral
\begin{equation}
    \frac{1}{L}\int_0^Ldz\int_0^Ldz'e^{-k|z-z'|}=\frac{2}{k}f(kL),
\end{equation}
where
\begin{equation}
    f(kL)=1-\frac{1-e^{-kL}}{kL},
\end{equation}
We have
\begin{equation}
    \bm{H}_D(\bm{k})=-4\pi\left(\frac{k_xk_y}{k^2}f(kL)m_y,\frac{k_y^2}{k^2}f(kL)m_y,[1-f(kL)]m_z\right).
\end{equation}
\begin{equation}
    E_{stat}=-\frac{1}{2}\bm{H}_D(\bm{k})\cdot\bm{m}=\frac{2\pi k_y^2}{k^2}f(kL)m_y^2+2\pi[1-f(kL)]m_z^2.
\end{equation}

\subsubsection{Total energy in k-space}
Combining the three components above and ignoring the constant part ($-2K_{x}-J$), the total energy is
\begin{equation}
E(\bm{k}) = \left(\frac{K_{x}+J}{2M_{s}^{2}}+\frac{2\pi k_{y}^{2}}{k^{2}}f(kL)\right)m_{y}^{2} + \left(\frac{K_{T}+J}{2M_{s}^{2}}+2\pi[1-f(kL)]\right)m_{z}^{2}+\frac{K_{x}}{2M_{s}^{2}}l_{y}^{2} + \frac{K_{T}}{2M_{s}^{2}}l_{z}^{2}.
\end{equation}
Define $\omega_{x} = \frac{2\gamma K_{x}}{M_{s}}$, $\omega_{T} = \frac{2\gamma K_{T}}{M_{s}}$, $\omega_{J} = \frac{2\gamma J}{M_{s}}$ and $\omega_{M} = 4\pi\gamma M_{s}$,
\begin{equation}
E(\bm{k}) = \left(\frac{\pi(\omega_{x}+\omega_{J})}{\omega_{M}}+\frac{2\pi k_{y}^{2}}{k^{2}}f(kL)\right)m_{y}^{2} + \left(\frac{\pi(\omega_{T}+\omega_{J})}{\omega_{M}}+2\pi[1-f(kL)]\right)m_{z}^{2}+\frac{\pi\omega_{x}}{\omega_{M}}l_{y}^{2} + \frac{\pi\omega_{T}}{\omega_{M}}l_{z}^{2}
\end{equation}

\subsubsection{Generalized Coordinate and Momentum}
We choose the generalized coordinate to be
\begin{equation}
\bm{x} = (m_{y},\ m_{z}).
\end{equation}
The generalized velocity is
\begin{equation}
\bm{v} = (-\omega_{T}l_{z},\ \omega_{x}l_{y}).
\end{equation}
The total energy (or Hamiltonian) is
\begin{equation}
\mathcal{H}=\left(\frac{\pi(\omega_{x}+\omega_{J})}{\omega_{M}}+\frac{2\pi k_{y}^{2}}{k^{2}}f(kL)\right)x_{1}^{2} + \left(\frac{\pi(\omega_{T}+\omega_{J})}{\omega_{M}}+2\pi[1-f(kL)]\right)x_{2}^{2}+\frac{1}{2}\frac{2\pi}{\omega_{T}\omega_{M}}v_{1}^{2} + \frac{1}{2}\frac{2\pi}{\omega_{x}\omega_{M}}v_{2}^{2}.
\end{equation}
The mass matrix $M$ is
\begin{equation}
M = \begin{bmatrix}
\frac{2\pi}{\omega_{T}\omega_{M}} & 0 \\
0 & \frac{2\pi}{\omega_{x}\omega_{M}} \\
\end{bmatrix}.
\end{equation}
The generalized momentum $\bm{p}$ is
\begin{equation}
\bm{p} = M\bm{v} = \left(-\frac{2\pi}{\omega_{M}}l_{z},\ \frac{2\pi}{\omega_{M}}l_{y}\right).
\end{equation}

Writing the Hamiltonian as a function of $\bm{x}$ and $\bm{p}$, and setting $\bm{x}$ and $\bm{p}$ real, we have
\begin{equation}
\mathcal{H}=\left(\frac{\pi(\omega_{x}+\omega_{J})}{\omega_{M}}+\frac{2\pi k_{y}^{2}}{k^{2}}f(kL)\right)x_{1}^{2} + \left(\frac{\pi(\omega_{T}+\omega_{J})}{\omega_{M}}+2\pi[1-f(kL)]\right)x_{2}^{2}+\frac{\omega_{T}\omega_{M}}{4\pi}p_{1}^{2} + \frac{\omega_{x}\omega_{M}}{4\pi}p_{2}^{2}.
\end{equation}

\subsubsection{Equations of Motion}
\label{sec:EOM}
Using Hamilton's equations, we can have equations of motion. We start from $\frac{d\bm{x}}{dt}$ to justify the definition of the generalized velocity.
\begin{equation}
\frac{d\bm{x}}{dt} = \frac{\partial\mathcal{H}}{\partial \bm{p}}
\end{equation}
\begin{equation}
\label{H1}
\frac{d}{dt}\begin{bmatrix}
x_{1} \\
x_{2} \\
\end{bmatrix} = \begin{bmatrix}
\frac{\omega_{T}\omega_{M}}{2\pi}p_{1} \\
\frac{\omega_{x}\omega_{M}}{2\pi}p_{2} \\
\end{bmatrix} = \begin{bmatrix}
-\omega_{T}l_{z} \\
\omega_{x}l_{y} \\
\end{bmatrix} = \bm{v}.
\end{equation}

The second Hamilton's equation gives
\begin{equation}
\frac{d\bm{p}}{dt} = -\frac{\partial \mathcal{H}}{\partial \bm{x}}
\end{equation}
\begin{equation}
\label{H2}
\begin{aligned}
\frac{d}{dt}\begin{bmatrix}
p_{1} \\
p_{2} \\
\end{bmatrix} &= -\begin{bmatrix}
\left(\frac{2\pi(\omega_{x}+\omega_{J})}{\omega_{M}}+\frac{4\pi k_{y}^{2}}{k^{2}}f(kL)\right)x_{1} \\
 \left(\frac{2\pi(\omega_{T}+\omega_{J})}{\omega_{M}}+4\pi[1-f(kL)]\right)x_{2} \\
\end{bmatrix} \\&= -\begin{bmatrix}
\frac{2\pi(\omega_{x}+\omega_{J})}{\omega_{M}}+\frac{4\pi k_{y}^{2}}{k^{2}}f(kL) & 0 \\
0 & \frac{2\pi(\omega_{T}+\omega_{J})}{\omega_{M}}+4\pi[1-f(kL)] \\
\end{bmatrix}\begin{bmatrix}
x_1 \\
x_2 \\
\end{bmatrix}
\end{aligned}
\end{equation}
Combining Eq.~\ref{H1} and \ref{H2}, we have
\begin{equation}
\frac{d^{2}}{dt^{2}}\begin{bmatrix}
x_{1} \\
x_{2} \\
\end{bmatrix} = M^{-1}\frac{d}{dt}\begin{bmatrix}
p_{1} \\
p_{2} \\
\end{bmatrix} = -\begin{bmatrix}
\omega_{T}(\omega_{x}+\omega_{J}) + \frac{2\omega_{T}\omega_{M} k_{y}^{2}}{k^{2}}f(kL) & 0 \\
0 & \omega_{x}(\omega_{T}+\omega_{J})+2\omega_{x}\omega_{M}[1-f(kL)] \\
\end{bmatrix}\begin{bmatrix}
x_1 \\
x_2 \\
\end{bmatrix}
\end{equation}
Using $\bm{x} = (m_{y},\ m_{z})$, we have
\begin{equation}
\label{EOM}
\frac{d^{2}}{dt^{2}}\begin{bmatrix}
m_{y} \\
m_{z} \\
\end{bmatrix} = -\begin{bmatrix}
\omega_{T}(\omega_{x}+\omega_{J}) + \frac{2\omega_{T}\omega_{M} k_{y}^{2}}{k^{2}}f(kL) & 0 \\
0 & \omega_{x}(\omega_{T}+\omega_{J})+2\omega_{x}\omega_{M}[1-f(kL)] \\
\end{bmatrix}\begin{bmatrix}
m_y \\
m_z \\
\end{bmatrix}
\end{equation}
Eq.~\ref{EOM} gives information on both magnon frequencies and normal modes. The magnon frequencies are
\begin{equation}
\begin{aligned}
    &\omega_y(\bm{k}) = \sqrt{\omega_{T}(\omega_{x}+\omega_{J}) + \frac{2\omega_{T}\omega_{M} k_{y}^{2}}{k^{2}}f(kL)},
    \\&\omega_z(\bm{k}) = \sqrt{\omega_{x}(\omega_{T}+\omega_{J})+2\omega_{x}\omega_{M}[1-f(kL)]},
\end{aligned}
\end{equation}
where $\omega_y$ corresponds to the high-frequency mode and $\omega_z$ is the low-frequency mode. From the expression, it is easy to tell that $\omega_y$ has an anisotropic dispersion and $\omega_z$ is isotropic, which is consistent with the result in our previous work. Eq.~\ref{EOM} also shows that the mode $\omega_y$ is associated with an oscillating $m_y$ and $\omega_z$ with an oscillating $m_z$, which is also consistent with all previous reports on magnons in easy-plane antiferromagnets.

\subsection{Dipolar field contributed by superconductors}
Now we turn to calculate the dipolar field contributed by the superconductor under the antiferromagnet. The thickness of the superconductor is $h$ and the superconductor is located at $-h<z<0$. 
\subsubsection{Dipolar field from magnons}
We first calculate the dipolar field from antiferromagnetic magnons in the region of $-h<z<0$.
\begin{equation}
    \bm{H}_{mag}(\bm{k},z)=-\int_0^Ldz'\Gamma(\bm{k},z-z')\bm{m}.
\end{equation}
Using the integral
\begin{equation}
    \int_0^Ldz'e^{-k|z-z'|}=e^{kz}\frac{1-e^{-kL}}{k},
\end{equation}
We have
\begin{equation}
\begin{aligned}
    &H_{mag,x} = \frac{2\pi k_x}{k} e^{kz}(1-e^{-kL})\left(im_z-\frac{k_y}{k}m_y\right),
    \\&H_{mag,y} = \frac{2\pi k_y}{k}e^{kz}(1-e^{-kL})\left(im_z-\frac{k_y}{k}m_y\right),
    \\&H_{mag,z} = 2\pi e^{kz}(1-e^{-kL})\left(m_z+i\frac{k_y}{k}m_y\right).
\end{aligned}
\end{equation}
In our experiment, the superconductor is BSCCO, which has a very anisotropic superfluid density. The superfluid density $\rho_{ab}$ within each CuO$_2$ layer is much larger than the superfluid density $\rho_c$ across layers. This means the current induced by $H_{mag,z}$ in BSCCO is much larger than that induced by $H_{mag,x}$ or $H_{mag,y}$. In the calculations below, We only consider the current within each CuO$_2$ layer.

\subsubsection{Self-consistent calculation for the magnetic field in the superconductor}
This part is the central task for calculating the coupling between the antiferromagnet and the superconductor. We use the same self-consistent equation in Ref.~\cite{borstObservationControlHybrid2023}. The total magnetic field $\bm{H}_{SC}$ in the superconductor has two contributions: one is the dipolar field $\bm{H}_{mag}$ from the magnons in the antiferromagnet, the other is the dipolar field $\bm{H}_{D,SC}$ from the Meissner current. However, the Meissner current depends on $\bm{H}_{SC}$. Therefore, we need to self-consistently solve for $\bm{H}_{SC}$ in the superconductor.

Using the Maxwell equation
\begin{equation}
    \nabla\times\bm{E} = -\frac{1}{c}\frac{\partial \bm{B}}{\partial t},
\end{equation}
We have
\begin{equation}
    \nabla\times\bm{E}=\frac{i\omega}{c}\bm{H}_{SC}.
\end{equation}
We used the relation $\bm{B}_{SC} = \bm{H}_{SC}$ because there is no magnetization in the superconductor. Using the relation $\bm{J} = \sigma\bm{E}$, we calculate the induced current in the superconductor.
\begin{equation}
    \nabla\times\bm{J} = \frac{i\omega\sigma}{c}\bm{H}_{SC}.
\end{equation}
We use the relation $\bm{J} = c\left(\nabla\times\bm{M}^{\textrm{eff}}\right)$ to derive the effective magnetization induced by the current in the superconductor.
\begin{equation}
    \nabla\times(\nabla\times\bm{M}^{\textrm{eff}})=\frac{i\omega\sigma}{c^2}\bm{H}_{SC}
\end{equation}
\begin{equation}
    \nabla(\nabla\cdot\bm{M}^{\textrm{eff}})-\nabla^2\bm{M}^{\textrm{eff}} = \frac{i\omega\sigma}{c^2}\bm{H}_{SC}
\end{equation}
In BSCCO, the superfluid density within each CuO$_2$ layer is significantly higher than that across CuO$_2$ layers, so we focus on the $z$-component of $\bm{H}_{SC}$.
\begin{equation}
    \frac{\partial}{\partial z}\left(ik_xM_x^{\textrm{eff}}+ik_yM_y^{\textrm{eff}}+\frac{\partial M_{z}^{\textrm{eff}}}{\partial z}\right)-\left(-k^2M_{z}^{\textrm{eff}}+\frac{\partial^2 M_{z}^{\textrm{eff}}}{\partial z^2}\right) = \frac{i\omega\sigma}{c^2}H_{SC,z}
\end{equation}
\begin{equation}
    \frac{ik_x}{k^2}\frac{\partial M_x^{\textrm{eff}}}{\partial z}+\frac{ik_y}{k^2}\frac{\partial M_y^{\textrm{eff}}}{\partial z}+M_{z}^{\textrm{eff}} = \frac{i\omega\sigma}{c^2k^2}H_{SC,z}
\end{equation}
The Meissner current induced dipolar field is given by
\begin{equation}
\begin{aligned}
H_{D,SC}^z = &-\int_{-h}^0dz'\left[\Gamma_{zx}(\bm{k},z-z')M_x^{\textrm{eff}}(z')+\Gamma_{zy}(\bm{k},z-z')M_y^{\textrm{eff}}(z')+\Gamma_{zz}(\bm{k},z-z')M_z^{\textrm{eff}}(z')\right]
\\=&-2\pi k\int_{-h}^0dz'\left[\sign(z-z')\frac{e^{-k|z-z'|}}{k}\left(ik_xM_{x}^{\textrm{eff}}+ik_yM_y^{\textrm{eff}}\right)-e^{-k|z-z'|}M_z^{\textrm{eff}}\right].
\end{aligned}
\end{equation}
Note that
\begin{equation}
    \frac{\partial}{\partial z'}e^{-k|z-z'|} = \sign(z-z')ke^{-k|z-z'|}.
\end{equation}
We have
\begin{equation}
    H_{D,SC}^z = -2\pi k\int_{-h}^0dz'\left[\left(\frac{\partial}{\partial z'}e^{-k|z-z'|}\right)\left(\frac{ik_x}{k^2}M_{x}^{\textrm{eff}}+\frac{ik_y}{k^2}M_y^{\textrm{eff}}\right)-e^{-k|z-z'|}M_z^{\textrm{eff}}\right].
\end{equation}
Integrating by parts, we have
\begin{equation}
\label{eq:relation_M_B}
\begin{aligned}
    H_{D,SC}^z=&-2\pi k\int_{-h}^0dz'\left[-e^{-k|z-z'|}\left(\frac{ik_x}{k^2}\frac{\partial M_{x}^{\textrm{eff}}}{\partial z'}+\frac{ik_y}{k^2}\frac{\partial M_{y}^{\textrm{eff}}}{\partial z'}\right)-e^{-k|z-z'|}M_z^{\textrm{eff}}\right]
    \\=&\ 2\pi k\int_{-h}^0dz'e^{-k|z-z'|}\left(\frac{ik_x}{k^2}\frac{\partial M_{x}^{\textrm{eff}}}{\partial z'}+\frac{ik_y}{k^2}\frac{\partial M_{y}^{\textrm{eff}}}{\partial z'}+M_z^{\textrm{eff}}\right)
    \\=&\ \frac{2\pi i\omega\sigma}{c^2k}\int_{-h}^0dz'e^{-k|z-z'|}H_{SC,z}(z').
\end{aligned}
\end{equation}
This expression is consistent with the physical picture that the dipolar field $\bm{H}_{D,SC}$ from the Meissner current has a dependency on the total magnetic field $\bm{H}_{SC}$ in the superconductor.
Now we can write the self-consistent equation for $H_{SC,z}$.
\begin{equation}
\label{eq:SCF}
\begin{aligned}
    H_{SC,z}(z) & = H_{D,SC}^z(z)+H_{mag,z}(z)
    \\&=\frac{2\pi i\omega\sigma}{c^2k}\int_{-h}^0dz'e^{-k|z-z'|}H_{SC,z}(z')+H_{mag,z}(z).
\end{aligned}
\end{equation}
Denote $H_{mag,z}(z)=H_0e^{kz}$, where $H_0=2\pi(1-e^{-kL})\left(m_z+i\frac{k_y}{k}m_y\right)$. We use a trial solution $H_{SC,z} = H_0(A_1e^{-\kappa z}+A_2e^{\kappa z})$ to solve Eq.~\ref{eq:SCF}.
\begin{equation}
    A_1e^{-\kappa z}+A_2e^{\kappa z} = \frac{2\pi i\omega\sigma}{c^2k}\int_{-h}^0dz'e^{-k|z-z'|}(A_1e^{-\kappa z'}+A_2e^{\kappa z'})+e^{kz}
\end{equation}
By equating the prefactors of terms $e^{\kappa z}$ and $e^{-\kappa z}$, we have
\begin{equation}
    \kappa = \sqrt{k^2-\frac{4\pi i\omega\sigma}{c^2}}.
\end{equation}
By equating the prefactors of terms $e^{kz}$ and $e^{-kz}$, we have
\begin{equation}
\begin{aligned}
    &A_1 =\frac{ic^2k^2}{2\pi\omega\sigma }\tilde{a}_1,
    \\&A_2=\frac{ic^2k^2}{2\pi\omega\sigma }\tilde{a}_2,
\end{aligned}
\end{equation}
where
\begin{equation}
\begin{aligned}
    &\tilde{a}_1 = \left(\frac{a_-^2}{a_+}\right)\frac{e^{-2\kappa h}}{1-\left(\frac{a_-}{a_+}\right)^2e^{-2\kappa h}},
    \\&\tilde{a}_2 = -\frac{a_-}{1-\left(\frac{a_-}{a_+}\right)^2e^{-2\kappa h}},
    \\&a_{\pm} = 1\pm\frac{\kappa}{k}.
\end{aligned}
\end{equation}
\subsubsection{Dipolar field from the circulating current}
Now we can calculate the dipolar field $\bm{H}_{D,SC}$ contributed by the Meissner current in the superconductor. We start with the $z$-component of the dipolar field in the antiferromagnet region ($0<z<L$). Using Eq.~\ref{eq:relation_M_B}, we have
\begin{equation}
\begin{aligned}
    H_{D,SC}^z(0<z<L)&=\frac{2\pi i\omega\sigma}{c^2k}\int_{-h}^0dz'e^{-k|z-z'|}H_{SC,z}(z')
    \\&=-kH_0\int_{-h}^0dz'e^{-k(z-z')}(\tilde{a}_1e^{-\kappa z'}+\tilde{a}_2e^{\kappa z'})
    \\&=-kH_0e^{-kz}\left(\tilde{a}_1\frac{1-e^{(\kappa-k)h}}{k-\kappa}+\tilde{a}_2\frac{1-e^{-(\kappa+k)h}}{k+\kappa}\right)
    \\&=-H_0e^{-kz}\left(\tilde{a}_1\frac{1-e^{(\kappa-k)h}}{a_-}+\tilde{a}_2\frac{1-e^{-(\kappa+k)h}}{a_+}\right)
    \\&=-H_0\frac{\frac{a_-}{a_+}e^{-kz}}{1-\left(\frac{a_-}{a_+}\right)^2e^{-2\kappa h}}\left(e^{-2\kappa h}-e^{-(\kappa+k)h}-1+e^{-(\kappa+k)h}\right)
    \\&=H_0 a_+ a_- e^{-kz}\frac{1-e^{-2\kappa h}}{a_+^2-a_-^2e^{-2\kappa h}}
\end{aligned}
\end{equation}
Note that
\begin{equation}
    a_+a_-=1-\left(\frac{\kappa}{k}\right)^2=\frac{4\pi i\omega\sigma}{c^2k^2}.
\end{equation}
We have
\begin{equation}
\begin{aligned}
    H_{D,SC}^z(0<z<L)&=H_0\frac{4\pi i\omega\sigma}{c^2k^2}e^{-kz}\frac{1-e^{-2\kappa h}}{a_+^2-a_-^2e^{-2\kappa h}}
    \\&=\frac{8\pi^2 i\omega\sigma}{c^2}\left(1-e^{-kL}\right)e^{-kz}\frac{1-e^{-2\kappa h}}{(k+\kappa)^2-(k-\kappa)^2e^{-2\kappa h}}\left(m_z+i\frac{k_y}{k}m_y\right).
\end{aligned}
\end{equation}
To calculate the $y$-component of $\bm{H}_{D,SC}$, we again use the dipolar tensor.
\begin{equation}
H_{D,SC}^y(0<z<L)=\left(\frac{ik_y}{k}\right)2\pi k\int_{-h}^0dz'\left[\frac{e^{-k|z-z'|}}{k}\left(ik_xM_{x}^{\textrm{eff}}+ik_yM_y^{\textrm{eff}}\right)-e^{-k|z-z'|}M_z^{\textrm{eff}}\right]
\end{equation}
Note that
\begin{equation}
    H_{D,SC}^z(0<z<L)=-2\pi k\int_{-h}^0dz'\left[\frac{e^{-k|z-z'|}}{k}\left(ik_xM_{x}^{\textrm{eff}}+ik_yM_y^{\textrm{eff}}\right)-e^{-k|z-z'|}M_z^{\textrm{eff}}\right],
\end{equation}
We get a simple relation between $H_{D,SC}^y$ and $H_{D,SC}^z$.
\begin{equation}
\begin{aligned}
    H_{D,SC}^y(0<z<L)&=-\frac{ik_y}{k}H_{D,SC}^z(0<z<L)
    \\&=\frac{8\pi^2 i\omega\sigma}{c^2}\left(1-e^{-kL}\right)e^{-kz}\frac{1-e^{-2\kappa h}}{(k+\kappa)^2-(k-\kappa)^2e^{-2\kappa h}}\left[-i\frac{k_y}{k}m_z+\left(\frac{k_y}{k}\right)^2m_y\right].
\end{aligned}
\end{equation}
Please note that this simple relation comes directly from the structure of the dipolar tensor, which originates from $\nabla\cdot\bm{B} = 0$.

\subsection{Dipolar magnon dispersion of antiferromagnets on a superconductor}
With the superconductor, there is an additional energy term $E_{SC}$ coming from the dipolar field contributed by the superconductor.
\begin{equation}
\begin{aligned}
    E_{SC} &= -\frac{1}{L}\int_{0}^{L}dz\bm{H}_{D,SC}(z)\cdot\bm{m}
    \\&=-\frac{1}{L}\int_{0}^{L}dz(H_{D,SC}^ym_y+H_{D,SC}^zm_z)
    \\&=-\frac{8\pi^2 i\omega\sigma}{c^2}\left(1-e^{-kL}\right)\frac{1-e^{-2\kappa h}}{(k+\kappa)^2-(k-\kappa)^2e^{-2\kappa h}}\frac{1}{L}\int_{0}^{L}dz e^{-kz}\left[\left(\frac{k_y}{k}\right)^2m_y^2+m_z^2\right]
    \\&=-\frac{8\pi^2 i\omega\sigma}{c^2}\frac{1-e^{-2\kappa h}}{(k+\kappa)^2-(k-\kappa)^2e^{-2\kappa h}}\frac{\left(1-e^{-kL}\right)^2}{kL}\left[\left(\frac{k_y}{k}\right)^2m_y^2+m_z^2\right]
    \\&=-\frac{8\pi^2 i\omega\sigma}{c^2}kL[1-f(kL)]^2\frac{1-e^{-2\kappa h}}{(k+\kappa)^2-(k-\kappa)^2e^{-2\kappa h}}\left[\left(\frac{k_y}{k}\right)^2m_y^2+m_z^2\right]
    \\&=-\frac{8\pi^2 i\omega\sigma}{c^2}\eta(k)\left[\left(\frac{k_y}{k}\right)^2m_y^2+m_z^2\right],
\end{aligned}
\end{equation}
where
\begin{equation}
    \eta(k)=kL[1-f(kL)]^2\frac{1-e^{-2\kappa h}}{(k+\kappa)^2-(k-\kappa)^2e^{-2\kappa h}}.
\end{equation}
In a superconductor, the imaginary part of $\sigma$ is determined by the London penetration depth $\lambda_L$.
\begin{equation}
    \sigma=\sigma_1+i\sigma_2,
\end{equation}
\begin{equation}
    \sigma_2=\frac{c^2}{4\pi\omega\lambda_L^2},
\end{equation}
where $e$ and $m_e$ are the charge and mass of the electron. Then $E_{SC}$ is
\begin{equation}
    E_{SC} = \left(-\frac{8\pi^2i\omega\sigma_1}{c^2}+\frac{2\pi}{\lambda_L^2}\right)\eta(k)\left[\left(\frac{k_y}{k}\right)^2m_y^2+m_z^2\right].
\end{equation}
Using the same approach as in Section ~\ref{sec:EOM}, we have the equation of motion below.
\begin{equation}
\frac{d^{2}}{dt^{2}}\begin{bmatrix}
m_{y} \\
m_{z} \\
\end{bmatrix} = -\begin{bmatrix}
\Omega_{yy} & \Omega_{yz}\\
\Omega_{zy} & \Omega_{zz} \\
\end{bmatrix}\begin{bmatrix}
m_y \\
m_z \\
\end{bmatrix},
\end{equation}
\begin{equation}
\begin{aligned}
    \Omega_{yy} &=  \omega_{T}(\omega_{x}+\omega_{J}) + \frac{2\omega_{T}\omega_{M} k_{y}^{2}}{k^{2}}f(kL) + 2\omega_T\omega_M\frac{\eta(k)}{\lambda_L^2}\left(\frac{k_y}{k}\right)^2-i\omega\frac{8\pi\omega_T\omega_M\sigma_1}{c^2}\eta(k)\left(\frac{k_y}{k}\right)^2,
    \\ \Omega_{zz} &= \omega_{x}(\omega_{T}+\omega_{J})+2\omega_{x}\omega_{M}[1-f(kL)]+ 2\omega_x\omega_M\frac{\eta(k)}{\lambda_L^2}-i\omega\frac{8\pi\omega_x\omega_M\sigma_1}{c^2}\eta(k),
    \\ \Omega_{yz} &= \Omega_{zy} = 0.
\end{aligned}
\end{equation}
Since there are no coupled terms between $m_y$ and $m_z$, we can write two independent differential equations using the relation $-i\omega\bm{m}=\frac{d}{dt}\bm{m}$.
\begin{equation}
    \frac{d^2}{dt^2}m_y+\frac{8\pi\omega_T\omega_M\sigma_1}{c^2}\eta(k)\left(\frac{k_y}{k}\right)^2\left(\frac{d}{dt} m_y\right)+\left[\omega_{T}(\omega_{x}+\omega_{J}) + \frac{2\omega_{T}\omega_{M} k_{y}^{2}}{k^{2}}f(kL)+2\omega_T\omega_M\frac{\eta(k)}{\lambda_L^2}\left(\frac{k_y}{k}\right)^2\right]m_y=0
\end{equation}
\begin{equation}
    \frac{d^2}{dt^2}m_z+\frac{8\pi\omega_x\omega_M\sigma_1}{c^2}\eta(k)\left(\frac{d}{dt} m_z\right)+\left[\omega_{x}(\omega_{T}+\omega_{J})+2\omega_{x}\omega_{M}[1-f(kL)]+2\omega_x\omega_M\frac{\eta(k)}{\lambda_L^2}\right]m_z=0
\end{equation}
The two differential equations above show that $\sigma_1$ only contributes to the damping of magnon modes, while $\sigma_2$ (or the London penetration depth $\lambda_L$) modifies the magnon dispersion. The magnon frequencies are
\begin{equation}
\label{eq:dispersion_relation_+}
    \omega_+(\bm{k})=\sqrt{\omega_{T}(\omega_{x}+\omega_{J}) + \frac{2\omega_{T}\omega_{M} k_{y}^{2}}{k^{2}}f(kL)+2\omega_T\omega_M\frac{\eta(k)}{\lambda_L^2}\left(\frac{k_y}{k}\right)^2},
\end{equation}
\begin{equation}
    \omega_-(\bm{k}) = \sqrt{\omega_{x}(\omega_{T}+\omega_{J})+2\omega_{x}\omega_{M}[1-f(kL)]+2\omega_x\omega_M\frac{\eta(k)}{\lambda_L^2}}.
\end{equation}

In experiments, we focus on the high-frequency mode $\omega_+(\bm{k})$. Eq.~\ref{eq:dispersion_relation_+} relies on multiple magnetic parameters $\omega_T$, $\omega_x$, $\omega_J$ and $\omega_M$. However, it is very difficult to measure the temperature dependence of each magnetic parameter. Below we show that the dispersion $\omega_{+}(k,\lambda_L)$ can be determined from the measured magnon frequency $\omega_0$ and the group velocity $v_0$ at $k=0$ in a bare CrSBr flake without the need of directly measuring magnetic parameters.

In a bare CrSBr, the magnon frequency $\omega_0$ at $k=0$ is $\sqrt{\omega_T(\omega_x+\omega_J)}$, and the group velocity $v_0$ at $k=0$ is given by
\begin{equation}
    v_0 = \frac{\partial\omega_{y}}{\partial k}\bigg|_{k=0}=\frac{\omega_T\omega_ML}{2\sqrt{\omega_T(\omega_x+\omega_J)}}.
\end{equation}
Combining $\omega_0$ and $v_0$, we have the relation
\begin{equation}
\omega_T\omega_M=\frac{2\omega_0v_0}{L}.
\end{equation}
Therefore,
\begin{equation}
    \omega_y(k) = \sqrt{\omega_{0}^2 + \frac{4\omega_0 v_0}{L}\left(\frac{k_y}{k}\right)^2f(kL)},
\end{equation}
\begin{equation}
\label{eq:dispersion_vg}
    \omega_{+}(k,\lambda_L) = \sqrt{\omega_y^2+\frac{4\omega_0 v_0}{L}\left(\frac{k_y}{k}\right)^2\frac{\eta(k,\lambda_L,h,L)}{\lambda_L^2}}.
\end{equation}

\section{Low-frequency mode}
\begin{figure}[H]
\centering
\includegraphics[width=0.75\columnwidth]{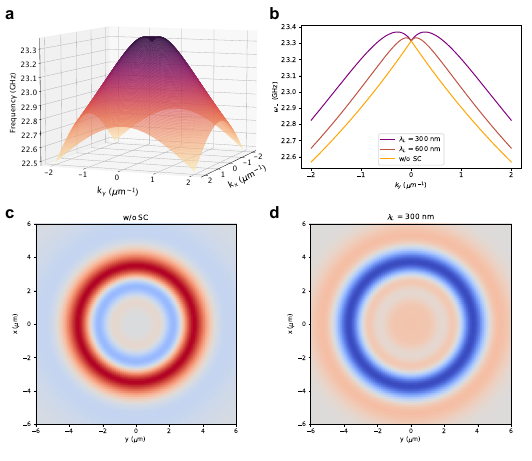}
\caption{(a) Calculated magnon frequency of the $\omega_-$ mode as a function of wavevector in the $k_x$-$k_y$ plane. (b) Magnon dispersion of the $\omega_-$ mode along the $k_y$ direction for different values of $\lambda_L$. (c,d) Simulated spatial distribution of the $\omega_-$ magnon mode at $\Delta t=1.77$ ns (c) in a bare CrSBr flake (thickness = 385 nm) and (d) in a CrSBr/superconductor heterostructure with $\lambda_L=300$ nm.}
\label{fig:low_freq_mode}
\end{figure}

In this section, we present the dispersion and propagation of the low frequency mode $\omega_{-}$. As shown in Fig.~\ref{fig:low_freq_mode}a, $\omega_{-}$ has an isotropic dispersion in the $k_x$-$k_y$ plane. The linecuts in Fig.~\ref{fig:low_freq_mode}b indicate that its dispersion is very nonlinear around $k=0$, where an M-shaped dispersion can be observed with $\lambda_L = $300 nm. However, the change in dispersion does not significantly influence the group velocity. Fig.~\ref{fig:low_freq_mode})(c,d) compare the spatial distribution of magnon wavepackets without and with a superconducting flake in proximity, which gives similar propagation range. This is the reason why we do not choose to use this mode to measure the superfluid density in BSCCO.

\section{Extracting group velocity from magnon dispersion}
As the dispersion is nonlinear even in the $k$-range covered by the optical excitation, we need to calculate the averaged group velocity $\langle v_g\rangle$.
\begin{equation}
    \langle v_g\rangle = \frac{d\langle x\rangle}{dt},
\end{equation}
where $\langle x\rangle$ is the expectation value of the magnon position. The probability function of the magnon distribution can be expressed with respect to the energy distribution. Noting that the total energy is proportional to $m_y^2$ or $m_z^2$, the probability of the energy distribution is
\begin{equation}
    P(x,t)=\frac{|m(x,t)|^2}{\int dx |m(x,t)|^2}.
\end{equation}
The expectation value of the magnon position is
\begin{equation}
\label{eq:ave_x}
\begin{aligned}
    \langle x(t)\rangle&=\int dx\ xP(x,t)
    \\&=\frac{\int dx\ x|m(x,t)|^2}{\int dx|m(x,t)|^2}.
\end{aligned}
\end{equation}
With
\begin{equation}
    m(x,t) = \int dk\ A(k)e^{i[kx-\omega(k)t]},
\end{equation}
where $A(k)$ is the $k$-distribution of the optical excitation at $t=0$. We first evaluate the denominator of Eq.~\ref{eq:ave_x}.
\begin{equation}
\begin{aligned}
    \int dx|m(x,t)|^2 &= \int dx \int dk \int dk' A(k) A^*(k')e^{i(k-k')x}e^{-i[\omega(k)-\omega(k')]t}
    \\&=2\pi\int dk\int dk' A(k)A^*(k')\delta(k-k')e^{-i[\omega(k)-\omega(k')]t}
    \\&=2\pi\int dk |A(k)|^2.
\end{aligned}
\end{equation}
Then we turn to the numerator.
\begin{equation}
    \int dx\ x|m(x,t)|^2=\int dx\int dk\int dk'\ xA(k)A^{*}(k')e^{i(k-k')x}e^{-i[\omega(k)-\omega(k')]t}.
\end{equation}
Noting that $xe^{i(k-k')x}=-i\frac{d}{dk}e^{i(k-k')x}$, we have
\begin{equation}
\begin{aligned}
    \int dx\ x|m(x,t)|^2&=-i\int dx\int dk\int dk'\ A(k)A^{*}(k')\frac{d[e^{i(k-k')x}]}{dk}e^{-i[\omega(k)-\omega(k')]t}
    \\&=-2\pi i\int dk\int dk'\ A(k)A^{*}(k')e^{-i[\omega(k)-\omega(k')]t}\frac{d}{dk}\delta(k-k').
\end{aligned}
\end{equation}
Integrating by parts, we have
\begin{equation}
    \begin{aligned}
    \int dx\ x|m(x,t)|^2&=2\pi i\int dk\int dk'\ \delta(k-k')\frac{d}{dk}\left[A(k)A^{*}(k')e^{-i[\omega(k)-\omega(k')]t}\right]
    \\&=2\pi i\int dk\int dk'\ \delta(k-k') \frac{dA(k)}{dk}A^*(k') e^{-i[\omega(k)-\omega(k')]t} 
    \\&+ 2\pi\int dk\int dk'\ \delta(k-k') \frac{d\omega(k)}{dk}t A(k)A^*(k') e^{-i[\omega(k)-\omega(k')]t}
    \\&=2\pi i\int dk\ \frac{dA(k)}{dk}A^*(k') + 2\pi\int dk\frac{d\omega(k)}{dk}t|A(k)|^2.
\end{aligned}
\end{equation}
Therefore we have the averaged group velocity
\begin{equation}
\label{eq:vg_extraction}
    \langle v_g\rangle=\frac{d\langle x\rangle}{dt} = \frac{\int dk\ |A(k)|^2 \frac{d\omega}{dk}}{\int dk\ |A(k)|^2}.
\end{equation}

\section{Criterion and accuracy of correlation-based measurements of the averaged group velocity}
The correlation-based measurements require taking two time traces $S_1(t)$ and $S_2(t)$ at two different spatial separations $\Delta x_1$ and $\Delta x_2$. $S_{1,2}(t)$ are given by
\begin{equation}
   S_{1,2}(t)=\int_{0}^{+\infty}dk \int_{-\infty}^{+\infty} dxP(x-\Delta x_{1,2})A(k)e^{i[kx-\omega(k)t]},
\end{equation}
where $A(k)$ is the k-distribution of the optical excitation and $P(x-\Delta x_{1,2})$ is the spatial distribution of the optical probe centered at $\Delta x_{1,2}$. Evaluating the spatial integral first, we have
\begin{equation}
\begin{aligned}
    S_{1,2}(t) &= \int_0^{+\infty} dk A(k)e^{i[k\Delta x_{1,2}-\omega(k)t]} \int_{-\infty}^{+\infty}dxP(x-\Delta x_{1,2})e^{ik(x-\Delta x_{1,2})}
    \\& = \int_0^{+\infty} dk A(k)P(k)e^{i[k\Delta x_{1,2}-\omega(k)t]},
\end{aligned}
\end{equation}
where $P(k)\equiv\int dxP(x-\Delta x_{1,2})e^{ik(x-\Delta x_{1,2})}$ is the Fourier transform of $P(x)$.

Now we evaluate the correlation function $C(\tau)$,
\begin{equation}
\begin{aligned}
    C(\tau) &= \int dt S_1^{*}(t)S_2(t+\tau)
    \\&=\int dt\int_{0}^{+\infty} dk\int_{0}^{+\infty} dk'A^{*}(k)P^{*}(k)A(k')P(k')e^{i(k\Delta x_2-k'\Delta x_1)}e^{-i[\omega(k)-\omega(k')]t}e^{-i\omega(k)\tau}.
\end{aligned}
\end{equation}
Note that
\begin{equation}
    \int dt e^{-i[\omega(k)-\omega(k')]t} = 2\pi\delta[\omega(k)-\omega(k')] = 2\pi\frac{\delta(k-k')}{|v_g(k)|}.
\end{equation}
Substituting this relation into the expression of $C(\tau)$, we have
\begin{equation}
\begin{aligned}
    C(\tau) &= 2\pi\int_0^{+\infty}dk\frac{|A(k)|^2|P(k)|^2}{|v_g(k)|}e^{i[k(\Delta x_2-\Delta x_1)-\omega(k)\tau]}
    \\&=2\pi\int_0^{+\infty}dk\frac{|A(k)|^2|P(k)|^2}{|v_g(k)|}e^{i[k\Delta s-\omega(k)\tau]}
    \\&=2\pi\int_0^{+\infty}dk|B(k)|^2e^{i[k\Delta s-\omega(k)\tau]},
\end{aligned}    
\end{equation}
where $\Delta s \equiv \Delta x_2 - \Delta x_1$ and $|B(k)|^2 = \frac{|A(k)|^2|P(k)|^2}{|v_g(k)|}$.

\begin{figure}[H]
\centering
\includegraphics[width=1.0\columnwidth]{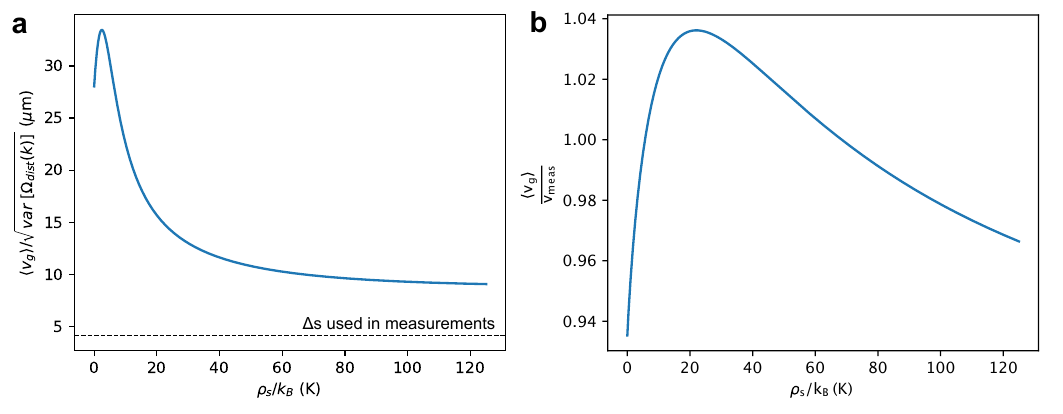}
\caption{(a) Comparison between $\Delta s$ used in measurements (the black dashed line) and the upper bound given by the derived criterion (the blue solid line). (b) Deviation between the measured $v_{meas} = \Delta s/\tau_{max}$ and $\langle v_g\rangle$.}
\label{fig:Criterion_Accuracy}
\end{figure}

Now we use a linear dispersion relation $\omega_l(k)$ with a slope of $\langle v_g\rangle$ to approximate $\omega(k)$,
\begin{equation}
    \omega_l(k) = \langle v_g\rangle(k-k_0)+\omega_0,
\end{equation}
\begin{equation}
    \omega(k) = \omega_l(k)+\Omega_{dist}(k),
\end{equation}
where $(k_0,\omega_0)$ is a random spot on the real dispersion relation $\omega(k)$ within the full width at half maximum of $|B(k)|^2$. $\Omega_{dist}(k)$ is the difference between $\omega_l(k)$ and $\omega(k)$, leading to wavepacket distortions. Later we will show why the choice of $(k_0,\omega_0)$ does not influence our analysis.
\begin{equation}
    C(\tau) = 2\pi\left(\int_0^{+\infty}dk|B(k)|^2e^{ik(\Delta s-\langle v_g\rangle\tau)}e^{-i\Omega_{dist}\tau}\right)e^{i(k_0\langle v_g\rangle-\omega_0)\tau},
\end{equation}
where $e^{i(k_0\langle v_g\rangle-\omega_0)\tau}$ is the carrier, and
\begin{equation}
    E(\tau) = \int_0^{+\infty}dk|B(k)|^2e^{ik(\Delta s-\langle v_g\rangle\tau)}e^{-i\Omega_{dist}(k)\tau}
\end{equation}
is the envelope. The amplitude of the envelope is given by
\begin{equation}
\label{eq:envelope}
    |E(\tau)|^2 = \int_0^{+\infty}\int_0^{+\infty}dk\ dk' |B(k)|^2|B(k')|^2e^{i(k-k')(\Delta s-\langle v_g\rangle\tau)}e^{i[\Omega_{dist}(k')-\Omega_{dist}(k)]\tau}.
\end{equation}
We expand $|E(\tau)|^2$ around $\tau=\tau_0 = \Delta s/\langle v_g\rangle$ to the second order,
\begin{equation}
    |E(\tau)|^2 = I_0+I_1^k+I_1^{\Omega}+I_2^k+I_2^{k\Omega}+I_2^\Omega
\end{equation}
\begin{equation}
\begin{aligned}
    &I_0 = \int_0^{+\infty}\int_0^{+\infty}dk\ dk' |B(k)|^2|B(k')|^2 e^{i[\Omega_{dist}(k')-\Omega_{dist}(k)]\tau_0}
    \\&I_1^k =- i\int_0^{+\infty}\int_0^{+\infty}dk\ dk' |B(k)|^2|B(k')|^2\langle v_g\rangle (k-k')e^{i[\Omega_{dist}(k')-\Omega_{dist}(k)]\tau_0}(\tau-\tau_0)
    \\&I_1^\Omega = i\int_0^{+\infty}\int_0^{+\infty}dk\ dk' |B(k)|^2|B(k')|^2[\Omega_{dist}(k')-\Omega_{dist}(k)]e^{i[\Omega_{dist}(k')-\Omega_{dist}(k)]\tau_0}(\tau-\tau_0)
    \\&I_2^{k} =-\frac{1}{2}\int_0^{+\infty}\int_0^{+\infty}dk\ dk' |B(k)|^2|B(k')|^2\langle v_g\rangle^2(k-k')^2e^{i[\Omega_{dist}(k')-\Omega_{dist}(k)]\tau_0}(\tau-\tau_0)^2
    \\&I_2^{k\Omega}=\int_0^{+\infty}\int_0^{+\infty}dk\ dk' |B(k)|^2|B(k')|^2\langle v_g\rangle(k-k')[\Omega_{dist}(k')-\Omega_{dist}(k)]e^{i[\Omega_{dist}(k')-\Omega_{dist}(k)]\tau_0}(\tau-\tau_0)^2
    \\&I_2^\Omega = -\frac{1}{2}\int_0^{+\infty}\int_0^{+\infty}dk\ dk' |B(k)|^2|B(k')|^2[\Omega_{dist}(k')-\Omega_{dist}(k)]^2e^{i[\Omega_{dist}(k')-\Omega_{dist}(k)]\tau_0}(\tau-\tau_0)^2.
\end{aligned}
\end{equation}
We note that all the terms above have a phase factor $e^{i[\Omega_{dist}(k')-\Omega_{dist}(k)]\tau_0}$. Below we show that Taylor expansion of this phase factor is a good approximation for all the terms above when $\Delta s$ is sufficiently small. There are only three integrals involved in the calculation.
\begin{equation}
    \Theta_0 = \int_0^{+\infty}\int_0^{+\infty}dk\ dk' |B(k)|^2|B(k')|^2 e^{i[\Omega_{dist}(k')-\Omega_{dist}(k)]\tau_0},
\end{equation}
\begin{equation}
    \Theta_1 = \int_0^{+\infty}\int_0^{+\infty}dk\ dk' |B(k)|^2|B(k')|^2 (k-k')e^{i[\Omega_{dist}(k')-\Omega_{dist}(k)]\tau_0},
\end{equation}
\begin{equation}
    \Theta_2 = \int_0^{+\infty}\int_0^{+\infty}dk\ dk' |B(k)|^2|B(k')|^2 (k-k')^2e^{i[\Omega_{dist}(k')-\Omega_{dist}(k)]\tau_0}.
\end{equation}
It is easy to show that
\begin{equation}
\begin{aligned}
    &I_0 = \Theta_0,
    \\&I_1^{k}\propto \Theta_1,\ I_1^{\Omega} \propto\frac{\partial\Theta_0}{\partial\tau_0},
    \\&I_2^k \propto\Theta_2,\ I_2^{k\Omega} \propto \frac{\partial\Theta_1}{\partial\tau_0},\ I_2^{\Omega} \propto \frac{\partial^2\Theta_0}{\partial\tau_0^2}.
\end{aligned}
\end{equation}
For $\Theta_0$, the first two nonzero terms are the zeroth and second-order terms $\Theta_0^{(0)}$ and $\Theta_0^{(2)}$. Denoting $N=\int_0^{+\infty}dk|B(k)|^2$, we have
\begin{equation}
    \Theta_0^{(0)} = N^2,
\end{equation}
\begin{equation}
\begin{aligned}
    \Theta_0^{(2)} &= \frac{1}{2}\int_0^{+\infty}\int_0^{+\infty}dk\ dk' |B(k)|^2|B(k')|^2[\Omega_{dist}(k')-\Omega_{dist}(k)]^2\tau_0^2
    \\&=\frac{\tau_0^2}{2}\int_0^{+\infty}\int_0^{+\infty}dk\ dk' |B(k)|^2|B(k')|^2[\Omega_{dist}(k')^2+\Omega_{dist}(k)^2]\\&-\tau_0^2\int_0^{+\infty}\int_0^{+\infty}dk\ dk' |B(k)|^2|B(k')|^2[\Omega_{dist}(k')\Omega_{dist}(k)]
    \\&=\tau_0^2N^2(\langle\Omega_{dist}(k)^2\rangle-\langle\Omega_{dist}(k)\rangle^2)=\tau_0^2N^2\textrm{var}[\Omega_{dist}(k)].
\end{aligned}
\end{equation}
To make the Taylor expansion a good approximation, it requires
\begin{equation}
    \frac{\Theta_0^{(2)}}{\Theta_0^{(0)}} = \tau_0^2\ \textrm{var}[\Omega_{dist}(k)] = \frac{\Delta s^2}{\langle v_g\rangle^2}\ \textrm{var}[\Omega_{dist}(k)] \ll 1,
\end{equation}
which gives the criterion
\begin{equation}
\label{eq:criterion}
    \boxed{\Delta s\ll\frac{\langle v_g\rangle}{\sqrt{\textrm{var}[\Omega_{dist}(k)]}}}
\end{equation}
Using similar procedures, one can get the criteria for $\Theta_1$ and $\Theta_2$. Denoting the standard deviation of $k$ and $\Omega_{dist}(k)$ as $\sigma_k$ and $\sigma_{\Omega}$, we have
\begin{equation}
    \frac{\Theta_1^{(3)}}{\Theta_1^{(1)}} = \frac{\frac{1}{6}\tau_0^3\sigma_k\sigma_\Omega^3}{\tau_0\sigma_k\sigma_\Omega} = \frac{1}{6}\tau_0^2\sigma_\Omega^2\ll1,
\end{equation}
\begin{equation}
    \frac{\Theta_2^{(2)}}{\Theta_2^{(0)}} = \frac{\frac{1}{2}\tau_0^2\sigma_k^2\sigma_\Omega^2}{\sigma_k^2} = \frac{1}{2}\tau_0^2\sigma_\Omega^2\ll1.
\end{equation}
The most strict requirement is still given by Eq.~\ref{eq:criterion}. As shown in Fig.~\ref{fig:Criterion_Accuracy}(a), our measurement condition satisfies the criterion in the whole range of phase stiffness.

Now we expand all the terms in $|E(\tau)|^2$ to the first order of $[\Omega_{dist}(k')-\Omega_{dist}(k)]\tau_0$
\begin{equation}
\begin{aligned}
    &I_0 = N^2
    \\&I_1^k=-2N^2 \langle v_g\rangle\textrm{cov}[k,\Omega_{dist}(k)]\tau_0(\tau-\tau_0)
    \\&I_1^{\Omega} = -2N^2\textrm{var}[\Omega_{dist}(k)]\tau_0(\tau-\tau_0)
    \\&I_2^k=-N^2\langle v_g\rangle^2\textrm{var}(k)(\tau-\tau_0)^2
    \\&I_2^{k\Omega} = -N^2\langle v_g\rangle\textrm{cov}[k,\Omega_{dist}(k)](\tau-\tau_0)^2
    \\&I_2^\Omega = 0
\end{aligned}
\end{equation}

Therefore, we can obtain $\tau_{max}$ by evaluating $\frac{d|E(\tau)|^2}{d\tau}=0$.
\begin{equation}
\tau_{max} = \tau_0 -\tau_0\frac{\langle v_g\rangle\textrm{cov}[k,\Omega_{dist}(k)]+\textrm{var}[\Omega_{dist}(k)]}{\langle v_g\rangle^2\textrm{var}(k)+\langle v_g\rangle\textrm{cov}[k,\Omega_{dist}(k)]}
\end{equation}
This expression tells us how wavepacket distortions influence the extraction of averaged group velocity. We can calculate the ratio between $\langle v_g\rangle$ and measured group velocity $v_{meas}$.
\begin{equation}
    \frac{\langle v_g\rangle}{v_{meas}}=\frac{\tau_{max}}{\tau_0} = 1-\frac{\langle v_g\rangle\textrm{cov}[k,\Omega_{dist}(k)]+\textrm{var}[\Omega_{dist}(k)]}{\langle v_g\rangle^2\textrm{var}(k)+\langle v_g\rangle\textrm{cov}[k,\Omega_{dist}(k)]}
\end{equation}
Fig.~\ref{fig:Criterion_Accuracy}(b) shows the ratio for the whole range of phase stiffness.

\section{Relation between the London penetration depth and the phase stiffness}
The gauge-invariant phase-gradient energy density in a superconductor is
\begin{equation}
    F_{\theta} = \frac{\rho_s}{2}\left(\nabla\theta-\frac{q}{\hbar c}\bm{A}\right)^2,
\end{equation}
where $q=2e$ for Cooper pairs and $\bm{A}$ is the vector potential. In 3D, $\rho_s^{3D}$ has the unit of energy per unit length, while in 2D, $\rho_s^{2D}$ has the unit of energy. We consider a layered superconductor with interlayer spacing $d$ between superconducting layers. In this case, $\rho_s^{3D} = \rho_s^{2D}/d$, where $\rho_s^{2D}$ is the phase stiffness of each superconducting layer.

In the layered superconductor, the current density is given by
\begin{equation}
\label{eq:J_from_F}
    \bm{J} = -c\frac{\partial F_{\theta}}{\partial \bm{A}} = \frac{q\rho_s^{3D}}{\hbar}\left(\nabla\theta-\frac{q}{\hbar c}\bm{A}\right).
\end{equation}
By taking the curl of both sides of Eq.~\ref{eq:J_from_F}, we have
\begin{equation}
    \nabla\times\bm{J} = -\frac{q^2\rho_s^{3D}}{\hbar^2c}\nabla\times\bm{A} = -\frac{q^2\rho_s^{3D}}{\hbar^2c}\bm{B}.
\end{equation}
On the other hand, the London equation gives
\begin{equation}
    \nabla\times\bm{J} = -\frac{c}{4\pi\lambda_L^2}\bm{B}.
\end{equation}
By equating the two coefficients above, we have
\begin{equation}
    \frac{q^2\rho_s^{3D}}{\hbar^2c} = \frac{c}{4\pi\lambda_L^2}.
\end{equation}
Thus
\begin{equation}
    \rho_s^{3D} = \frac{\hbar^2c^2}{4\pi q^2\lambda_L^2} = \frac{\hbar^2 c^2}{16\pi e^2\lambda_L^2}.
\end{equation}
We further convert $\rho_s^{3D}$ to the phase stiffness $\rho_s^{2D}$ for each superconducting layer.
\begin{equation}
\label{eq:conversion_lambdaL_rhos}
    \rho_s^{2D} = \rho_s^{3D}d = \frac{\hbar^2 c^2d}{16\pi e^2\lambda_L^2}
\end{equation}

\section{Procedure of measuring the phase stiffness energy}
This section describes all details on how to measure the phase stiffness energy based on magnon propagation.

\subsection{Experimental protocol}
\label{subsec:protocol}
As introduced in the main text, we use stripe-shaped pump and probe beams to measure magnon propagation along the $a$-axis of CrSBr. Along the magnon propagation direction, the full width at half maximum (FWHM) of both pump and probe beams is 1.4 $\mu$m. The pump wavelength is 520 nm and the probe wavelength is 700 nm. The spatial separation between the pump and probe beams is programmably controlled by the combination of a pair of galvo mirrors and a 4$f$ imaging system. At each temperature and magnetic field, we follow the experimental protocol described below to take time traces of transient reflectance at two different spatial separations. We will use the data taken on the CrSBr/BSCCO sample without hBN spacer at $T$ = 59 K and $B_a$ = 0.3 T as an example when introducing the protocol.
\begin{enumerate}
\item Determine the two spatial separations $\Delta x_{1,2}$ for the measurement. There are two standards for the determination: (i) $\Delta x > \sqrt{2}D$, where $D$ is the beam diameter of both pump and probe foci. This standard helps minimize the pump-probe signals from the direct overlap of pump and probe beams. (ii) $\Delta x$ should meet the criterion in Eq.~\ref{eq:criterion}. At $T$ = 59 K and $B_a$ = 0.3 T, we choose $\Delta x_1$ = 5.50 $\mu$m and $\Delta x_2$ = 9.63 $\mu$m, which correspond to the galvo mirror voltages $V_1$ = 40 mV and $V_2$ = 70 mV, respectively.
\item Set the galvo mirror voltage to $V_1-\Delta V$ ($\Delta V$ = 10 mV in this case), and then $V_1$. Take one transient reflectance time trace at $\Delta x_1$. This procedure is to avoid the hysteresis of the galvo mirror.
\item Set the galvo mirror voltage to $V_2$. Take one transient reflectance time trace at $\Delta x_2$.
\item Repeat Steps 2 and 3 for 25 iterations. The alternating measurements at $\Delta x_1$ and $\Delta x_2$ average out the systematic error associated with the pointing direction of the galvo mirror.
\end{enumerate}
Following the protocol above, we get two time traces $S_1(t)$ and $S_2(t)$ in Fig.~\ref{fig:determine_T_theta}a. We then calculate the correlation function $C(\tau) = \langle S_1(t)S_2(t+\tau)\rangle_t$ and find the time delay $\tau_{max}$ = 0.541 ns at which the envelope of $C(\tau)$ is maximized (Fig.~\ref{fig:determine_T_theta}b). Therefore, $\langle v_g\rangle = \frac{\Delta x_2-\Delta x_1}{\tau_{max}}$ = 7.62 km/s.

\subsection{Determination of the magnon dispersion}
Since we study the one-dimensional (1D) propagation along the $y$-axis ($a$-axis of CrSBr), we only need to consider the linecut $k_x = 0, k_y = k$. The 1D dispersion is
\begin{equation}
\label{eq:dispersion_vg}
    \omega_{+}(k,\lambda_L) = \sqrt{\omega_{0}^2 + \frac{4\omega_0 v_0}{L}f(kL)+\frac{4\omega_0 v_0}{L}\frac{\eta(k,\lambda_L,h,L)}{\lambda_L^2}}.
\end{equation}
Eq.~\ref{eq:dispersion_vg} shows that the magnon dispersion can be uniquely determined by $\lambda_L$, $\omega_0$, $v_0$, $L$ and $h$. The CrSBr thickness $L$ and BSCCO thickness $h$ can be directly measured by atomic force microscopy. As mentioned in the main text, $L$ = 385 nm and $h$ = 187 nm. $\omega_0$ can be extracted from the Fourier transform of the time trace. At $T$ = 59 K and $B_a$ = 0.3 T, $\omega_0 = 2\pi\times26$ GHz. 

$v_0$ can be determined by measuring the averaged group velocity $\langle v_{g0}\rangle$ of the magnon wavepacket on the bare CrSBr flake. Following the protocol described in Section \ref{subsec:protocol}, we get $\langle v_{g0}\rangle$ = 5.90 km/s. Using Eq.~\ref{eq:vg_extraction}, we obtain the relation
\begin{equation}
\label{eq:determine_v0}
  \frac{\int dk\ |A(k)|^2 \frac{d\omega_y(k,v0)}{dk}}{\int dk\ |A(k)|^2} = \langle v_{g0}\rangle,\ A(k) = e^{-k^2\sigma^2/2},  
\end{equation}
where $\sigma = \frac{D}{2\sqrt{\ln 2}}$ and $D = 1.4$ $\mu$m is the FWHM of the pump beam. Solving Eq.~\ref{eq:determine_v0} gives $v_0$ = 7.58 km/s.

\subsection{Conversion from group velocity to phase stiffness energy}
\begin{figure}[H]
\centering
\includegraphics[width=0.9\columnwidth]{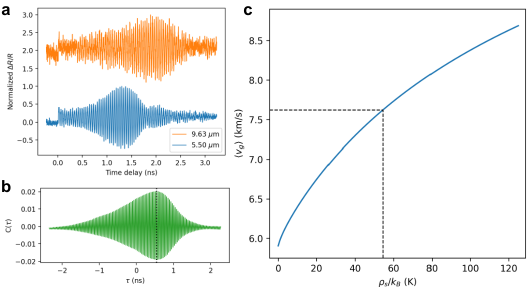}
\caption{(a) Example transient reflectance data measured for pump–probe separations at 5.50 and 9.63 $\mu$m in CrSBr/BSCCO at $T$ = 59 K and $B_a$ = 0.3 T. (b) The correlation function $C(\tau)$ calculated from the data in (a). The envelope of $C(\tau)$ peaks at $\tau = $ 0.541 ns indicated by the dashed line. (c) The relation between the averaged group velocity $\langle v_g\rangle$ and the phase stiffness $\rho_s$. The dashed lines mark the correspondence between $\langle v_g\rangle = 7.62$ km/s and $\rho_s/k_B = 54.5$ K.}
\label{fig:determine_T_theta}
\end{figure}

Using Eq.~\ref{eq:conversion_lambdaL_rhos}, the magnon dispersion can now be expressed as $\omega_+(k,\rho_s)$. In BSCCO, the interlayer spacing $d$ of CuO$_2$ planes is 1.54 nm. Fig.~\ref{fig:determine_T_theta}c shows the relation between the magnon wavepacket group velocity $\langle v_g\rangle$ and $\rho_s$ calculated from Eq.~\ref{eq:vg_extraction}. Since $\langle v_g\rangle$ is a single-valued function of $\rho_s$, the measured $\langle v_g\rangle$ = 7.62 km/s uniquely corresponds to $\rho_s/k_B$ = 54.5 K.

\bibstyle{apsrev4-2}

\bibliography{SI_text_citation}